\documentclass[journal,draftclsnofoot,onecolumn,12pt,twoside]{IEEEtranTCOM}

\usepackage{indentfirst}
\usepackage{multirow}
\usepackage{graphicx}
\usepackage{color}
\usepackage{xcolor}
\usepackage{listings}
\usepackage{enumerate}
\usepackage{subfigure}
\usepackage{amsmath,amsthm,amscd,amssymb,latexsym,upref,stmaryrd}
\usepackage[noend]{algpseudocode}
\usepackage{algorithmicx,algorithm}
\usepackage{amsfonts,hyperref,epsfig}
\usepackage{CJK}
\usepackage{float}
\usepackage{cite}
\usepackage{subfigure}
\usepackage{lipsum}
\usepackage{multicol}
\usepackage{setspace}

\renewcommand{\baselinestretch}{1.49}

\begin{document}

\title{Beamspace Channel Estimation in mmWave Systems via Cosparse Image Reconstruction Technique }

\author{Jie Yang,
Chao-Kai Wen,~\IEEEmembership{Member,~IEEE},
Shi Jin,~\IEEEmembership{Member,~IEEE},
\\and Feifei Gao,~\IEEEmembership{Member,~IEEE}
\thanks{ Jie Yang and Shi Jin are with the National Mobile Communications Research Laboratory, Southeast University, Nanjing, 210096, P. R. China (e-mail: yangjie@seu.edu.cn; jinshi@seu.edu.cn).}
\thanks{ Chao-Kai Wen is with the Institute of Communications Engineering, National Sun Yat-sen University, Kaohsiung, 804, Taiwan (e-mail: chaokai.wen@mail.nsysu.edu.tw).}
\thanks{ Feifei Gao is with Tsinghua National Laboratory for Information Science and Technology (TNList), Beijing, 100084, P. R. China (e-mail:  feifeigao@ieee.org).}
}

\markboth{IEEE Transactions on Communications}%
{Submitted paper}

\maketitle

\vspace{-5mm}
\begin{abstract}

This paper considers the beamspace channel estimation problem in 3D lens antenna array under a millimeter-wave communication system. We analyze the focusing capability of the 3D lens antenna array and the sparsity of the beamspace channel response matrix. Considering the analysis, we observe that the channel matrix can be treated as a 2D natural image; that is,  the channel is sparse, and the changes between adjacent
elements are subtle. Thus, for the channel estimation, we incorporate an image
reconstruction technique called \b{s}parse non-informative parameter estimator-based \b{c}osparse analysis \b{A}\b{M}\b{P} for \b{i}maging (SCAMPI) algorithm.
The SCAMPI algorithm is faster and more accurate than earlier algorithms such as orthogonal matching pursuit and support detection algorithms.
To further improve the SCAMPI algorithm, we model the channel distribution as a generic Gaussian mixture (GM) probability and embed the expectation maximization learning algorithm into the SCAMPI algorithm to learn the parameters in the GM probability.
We show that the GM probability outperforms the common uniform distribution used in image reconstruction. We also propose a phase-shifter-reduced selection network structure to decrease the power consumption of the system and prove that the SCAMPI algorithm is robust even if the number of phase shifters is reduced by 10$\%$.

\end{abstract}



\section{Introduction}

A millimeter-wave (mmWave) communication system plays a promising role in future fifth generation cellular networks. The mmWave band can offer larger bandwidth communication channels than currently used bands in commercial wireless systems, however, the penetration losses are larger on the mmWave than on the lower-frequency wave\cite{mmWave1,mmWave2,mmWave3,overview}.
Therefore, a large antenna array with highly directional transmission/reception should be made to compensate for the high penetration losses \cite{mmWaveMIMO1,mmWaveMIMO2}.
However, hardware complexity and power consumption are large because of the use of large antenna arrays \cite{hard1}.
Several architectures have been proposed to solve the hardware constraints in mmWave communication, including the hybrid analog/digital precoding combining architecture using phase array or lens \cite{hy0,hy1,hy2,hy3} and the low-resolution ADC architecture\cite{1bit1}. 

Among these architectures, the lens antenna array is one of the most interesting architectures because it has many advantages over the common antenna array \cite{lens1,lens2}. In particular, the lens can operate at very short pulse lengths and scan wider beamwidths than any previously known device\cite{lens1}; furthermore, the lens is capable of forming low sidelobe beams \cite{lens2}. Given these advantages, the lens antenna array was introduced into mmWave communication systems as transmit/receive front ends in \cite{lens3,lens4,lens5}. Implanting the lens antenna array into hybrid analog/digital precoding combining architecture also showed good promise \cite{beamspace,1,7}. \cite{beamspace} introduced the concept of beamspace channel model and the lens antenna array-based architecture of continuous aperture-phased (CAP)-MIMO transceiver. Since the CAP-MIMO transceiver can obtain signals directly in the beamspace through an array of feed antennas arranged on the focal surface of the lens, transceiver complexity can be reduced dramatically\cite{beamspace}. Recently, \cite{1} claimed that the array response of the 2D lens antenna array at the receiver/transmitter follows a ``sinc" function. Furthermore, the 3D lens antenna array response was derived from \cite{7}, which was given by the product of two ``sinc" functions.

Studies on \cite{beamspace,1,7} focused on the transmission architecture for lens-based mmWave systems.
Under the 3D lens antenna array setup, the mmWave lens system has both azimuth and elevation angle resolution capabilities. The ``sinc" type function infers that for a given angle of arrival (AoA)/departure (AoD) of the received/transmitted signal, only those antennas located near the energy focusing point receive/transmit significant power. We call the aforementioned phenomenon energy-focusing capability of the EM lens.
The energy-focusing capability of the EM lens antenna array, together with the multi-path sparsity of the mmWave channel, bring new techniques for mmWave communication systems, especially in channel estimation techniques.

\begin{figure} [htbp!]
\centering
\includegraphics[scale=0.38,bb= 0 0 300 340]{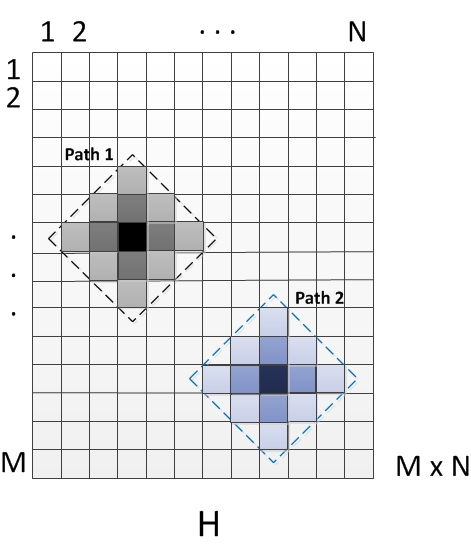}
\caption{Illustration the amplitude of channel matrix. We assume that two paths exist, and the effective entries are concentrated in squares in channel matrix painted with colors. Except for the values in these entries, the values of other white entries are nearly zero.}\label{sparsity}
\end{figure}

In this paper, we focus on the channel estimation in 3D lens-based mmWave systems.
Channel estimation problems for the lens mmWave system is challenging, especially when the antenna array is large and the number of RF chains is limited.
Several works utilize CS techniques to solve the beamspace channel estimation problem in the mmWave MIMO system \cite{111,LASSO,OMP,cs1,cs2,cs3,cs4,cs5}, where CS techniques bring numerous benefits. \cite{111} revealed that by using CS channel estimation algorithm, the system operation required a relatively smaller training overhead than the systems using exhaustive narrow beam scanning.
To adapt to various channel estimation scenarios, several CS algorithms have been introduced. Algorithms such as LASSO \cite{LASSO} and orthogonal matching pursuit (OMP) \cite{OMP} were introduced by \cite{111} for single-user channel estimation \cite{cs1}. Then, \cite{cs2,cs3,cs4,cs5,jin} considered multi-user systems. A joint OMP recovery algorithm performed at BS was proposed by \cite{cs2}. \cite{cs4} claimed that quadratic semi-defined programming algorithm outperforms existing algorithms in estimation error performance or training transmit power.
However, the aforementioned algorithms are not optimal for lens-based mmWave systems because the lens antenna array has energy-focusing capability, and the received signal matrix from the lens antenna array is characterized by sparsity and concentration \cite{1,7}, as shown in Fig. \ref{sparsity}. Thus, the channel estimation algorithm complexity can be simplified and the performance can be improved further.
For example, \cite{3} proposed a reliable support detection (SD)-based channel estimation scheme for mmWave lens systems, which performed better than the OMP algorithm. However, the SD algorithm only utilized the sparsity feature of the channel, and did not consider the clustering feature of the paths.

In this study, we consider the clustering feature of the paths, and design high-performance channel estimation algorithms based on image reconstruction techniques.
Our design leverages a simple observation: the channel of the 3D lens antenna array system can be seen as a 2D matrix, and the 2D channel matrix can be treated as a 2D natural image.
To obtain an idea regarding this concept, we randomly generate a channel with two paths.
The corresponding pseudo-color picture of the amplitude of the 2D channel matrix is depicted in Fig. \ref{sparsity}. The entries with stronger amplitude are painted with a relatively darker color.
We obtain two observations: 1) most entries of the channel matrix can be ignored as zero, and 2) the changes between most of the adjacent elements in the channel
matrix are subtle.

Inspired by these two observations, we employ a method called SCAMPI to the channel estimation. ``SCAMPI" is the abbreviation of ``\b{s}parse non-informative parameter estimator-based \b{c}osparse analysis \b{a}pproximate \b{m}essage-\b{p}assing (AMP) for \b{i}maging", which was used in compressive image reconstruction \cite{5}. Our design builds on the insights learned from image reconstruction, but it is the first to enable channel estimation, and at an accuracy suitable for 3D lens antenna array in the mmWave communication system. In Section III, we explain how we formulate the system model that incorporates SCAMPI to estimate the channel.
We further embed SCAMPI into the system that overcomes additional practical challenges. In particular, to perform SCAMPI,
the exact prior distribution of the channel and the noise variance are required. To address the issues, we model the channel distribution as a generic $L$-term Gaussian mixture (GM) probability distribution as that in \cite{em3} and use expectation-maximization (EM) algorithm to learn the GM parameters \cite{6,em3,em1,em2}.

The contributions of this paper are as follows:
\begin{itemize}
  \item We formulate the channel estimation problem in the 3D lens antenna array that can incorporate the SCAMPI algorithm to estimate the channel. We show that the SCAMPI algorithm can perform more effectively than other existing algorithms. To the best of our knowledge, this paper is the first study that demonstrates the channel-estimation bridging to an image reconstruction technique.

  \item The value of each pixel in an image basically obeys uniform distribution. However, we observe that the channel responses follow nearly sparse Gaussian distribution in real life. Therefore, we replace the uniform prior distribution of channel responses with the sparse Gaussian distribution. The replacement introduces a practical challenge because several unknown parameters (such as sparsity rate, mean, and variance) appear in sparse Gaussian priori probability distribution function. Thus, we introduce the EM learning algorithm to find the (locally) maximum likelihood estimates of the parameters. The EM learning of the parameters is then deduced. The result reveals that, with the help of the EM learning algorithm, the sparse Gaussian is more suitable to be priori probability distribution than the uniform distribution used in images.

  \item We propose a new phase-shifter reduced measurement matrix structure in which a random part of the phase shifter is disconnected from the entire network. The new structure can reduce the power consumption of the system. We analyze the effect of the measurement matrix structure on the performance of the SCAMPI algorithm. The simulation results show that the SCAMPI algorithm is robust even if the number of phase shifters is reduced by 10$\%$.
\end{itemize}

The rest of this paper is organized as follows. In Section \uppercase\expandafter{\romannumeral2}, we derive a 3D lens antenna array-based mmWave system model and introduce a new phase-shifter-reduced selection network structure. The SCAMPI algorithm is introduced in Section \uppercase\expandafter{\romannumeral3}, and the priori probability distributions of channel responses are also discussed. In Section \uppercase\expandafter{\romannumeral4}, we embed the EM learning algorithm into the SCAMPI algorithm and deduce the update expression of parameters in each iteration. The simulation results are discussed and compared in Section \uppercase\expandafter{\romannumeral5}. Finally, we conclude the paper in Section \uppercase\expandafter{\romannumeral6}.

{\bf Notations}---Throughout this paper, uppercase boldface $\mathbf{A}$ and lowercase boldface $\mathbf{a}$ denote matrices and vectors, respectively. For any matrix $\mathbf{A}$, the superscripts $\mathbf{A}^{T}$ and $\mathbf{A}^{H}$ stand for the transpose and conjugate-transpose, respectively, and $a_{m,n}$ denotes the $(m,n)$th element in matrix $\mathbf{A}$. If $\mathbf{A}$ is a non-singular square matrix, its matrix inverse is denoted as $\mathbf{A}^{-1}$. An identity matrix is denoted by $\mathbf{I}$ or $\mathbf{I}_{N}$ if it is necessary to specify its dimension $N$. For a vector $\mathbf{a}$, the 2-norm is denoted by $\|\mathbf{a}\|_{2}$.
For a set $\mathbb{E}$, $|\mathbb{E}|$ represents the number of elements in set $\mathbb{E}$. $\delta(\cdot)$ denotes the Dirac delta function, and ${\rm sinc}(x)=\frac{\sin(\pi x)}{\pi x}$ denotes the ``sinc" function. For a Gaussian random vector $\mathbf{z}$, $\mathbf{z}\sim \mathcal{N}(u,\sigma^2)$ denotes the probability distribution function for $\mathbf{z}$ with mean $u$ and variance $\sigma^2$. In addition, the expectation operators are denoted by $E\{\cdot\}$. $\lfloor a\rfloor$ denotes the largest integer no greater than $a$, and $\lceil a\rceil$ denotes the smallest integer no smaller than $a$.

\section{System Model}
\begin{figure}[htbp!]
\centering
\includegraphics[scale=0.37,bb= 0 0 750 340]{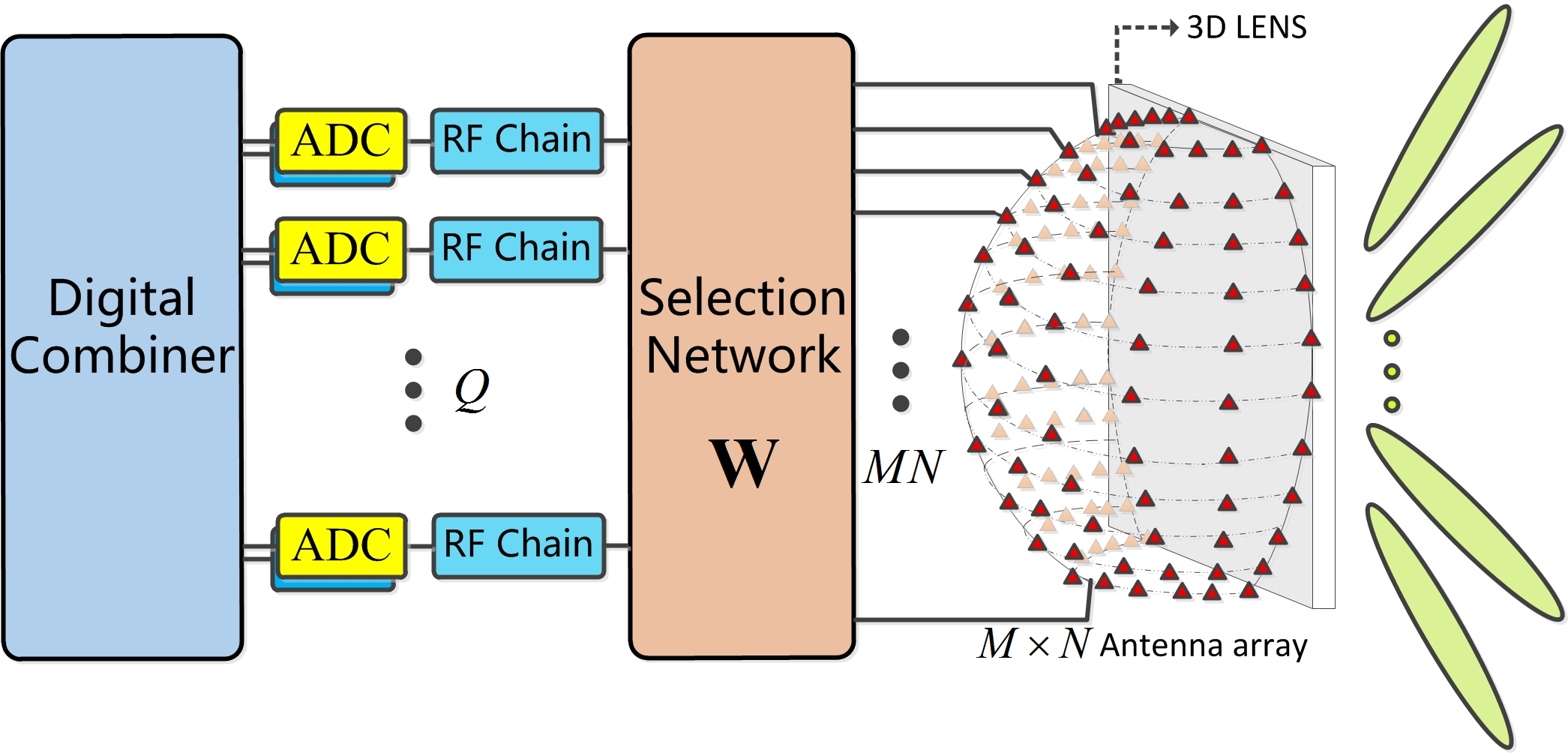}
\caption{Schematic of base station in 3D lens antenna array-based mmWave system, where the base station has an EM lens and an $M\times N$ antenna array placed on the focal plane of the lens.}\label{concentrated}
\end{figure}

\begin{figure}
\centering
\subfigure[Conventional selection network, which is composed of fully connected 1-bit phase shifters.]
 {
 \label{ps:a}
 \centering
 \includegraphics[scale=0.28,bb= 0 0 560 340]{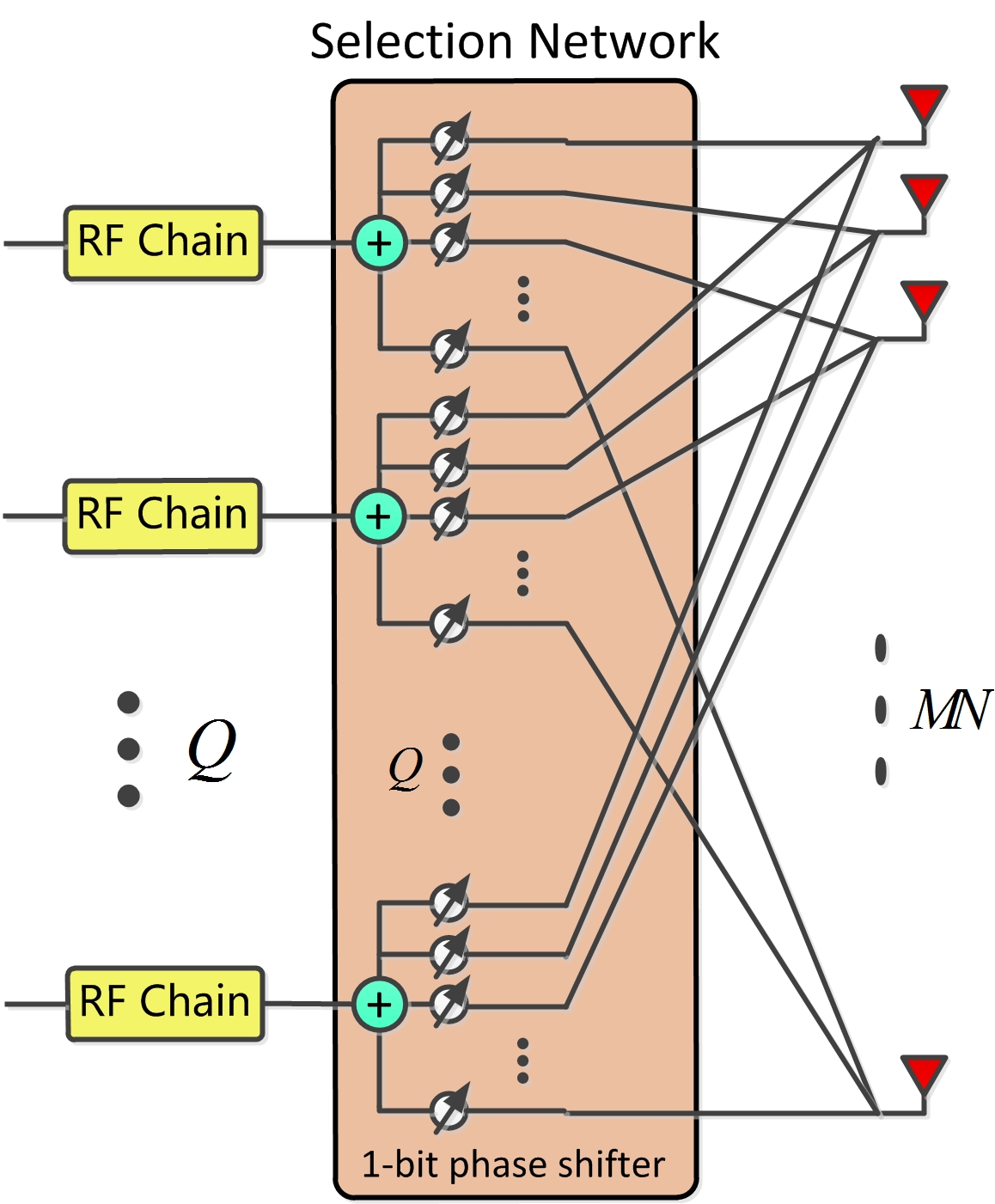}
 }
\hspace{0.35in}
\subfigure[Phase shifter reduced adaptive selection network in which a random part of the phase shifter is switched off.]{
 \label{ps:b}
 \centering
 \includegraphics[scale=0.28,bb= 0 0 560 370]{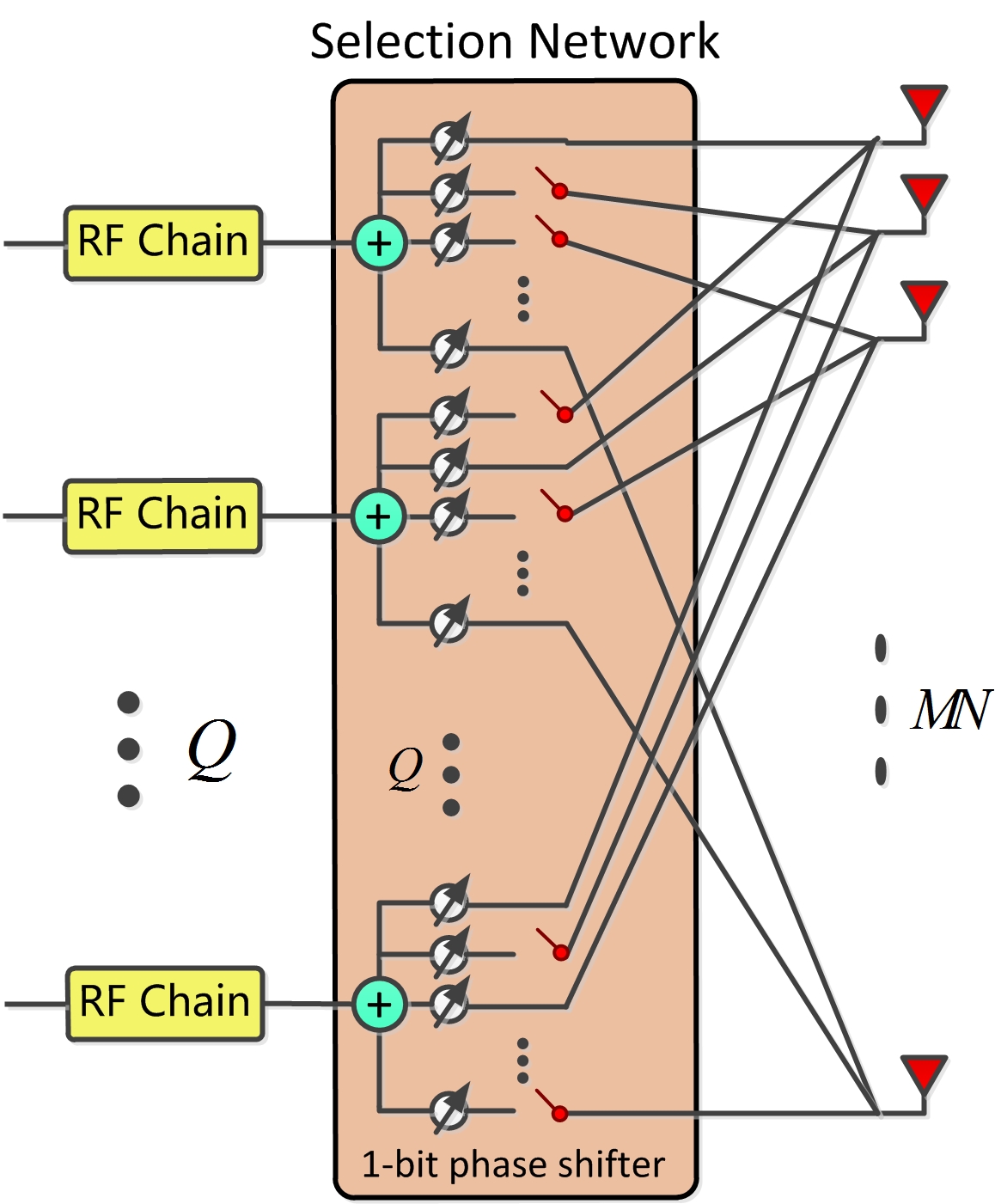}}
\caption{Schematic of two types of selection network with 1-bit phase shifters} \label{ps}
 \end{figure}

We employ a 3D mmWave lens antenna array, which has both azimuth and elevation angle-resolution capabilities. As illustrated in Fig. \ref{concentrated}, the BS has one lens equipped with a ${M\times N}$ antenna array, and the $MN$ antennas are connected to the $Q$ RF chains through the ${Q\times MN}$ selection network. In contrast to the conventional selection network, which uses the fully connected phase shifters (Fig. \ref{ps:a}), a phase shifter-reduced adaptive selection network (Fig. \ref{ps:b}) is proposed, in which some phase shifters are switched off. Specifically, the selection network can be expressed by a ${Q\times MN}$ matrix $\mathbf{W}$ with each entry being either $0$ or {$\pm 1$}.
The advantage of this architecture is that it can be easily configured for beam selection in the data transmission phase and for combiner in the channel estimation phase.


For ease of expression, we consider the system model with a single user. The system model can be easily extended to deal with the case with multiple users as long as the pilot signals for different users are orthogonal in time. In the uplink training phase, the user sends training symbol $s$ to the BS, and
then the received signal vector at the BS is given by
\begin{equation}\label{m1}
\mathbf{y}=\mathbf{h}s+\mathbf{n},
\end{equation}
where $\mathbf{n}\sim \mathcal{N}(0,\Delta)$ is a Gaussian noise vector, and $\mathbf{h} \in \mathbb{C}^{MN\times 1}$ is the beamspace channel vector.

Before proceeding, we first specify the beamspace channel vector $\mathbf{h}$ in the lens antenna array system. To this end, we recall
the 2D Saleh-Valenzuela channel model
\begin{equation}\label{nn}
\mathbf{h}=\sqrt{\frac{N_{a}}{L+1}}\sum_{l=0}^{L}\alpha^{(l)}\mathbf{a}(\phi_{y}^{(l)},\phi_{z}^{(l)}),
\end{equation}
where $N_{a}$ is the number of antennas, $L$ is the number of paths, and
$\alpha^{(l)}$ is the complex path gain of the $l$th path. In addition, $\phi_{y}^{(l)}$ and $\phi_{z}^{(l)}$ are azimuth and elevation AoAs of the incident plane wave, respectively, and $\mathbf{a}(\phi_{y}^{(l)},\phi_{z}^{(l)})$ is the antenna array response vector.

\begin{figure} 
\centering
\includegraphics[scale=0.38,bb= 0 0 477 450]{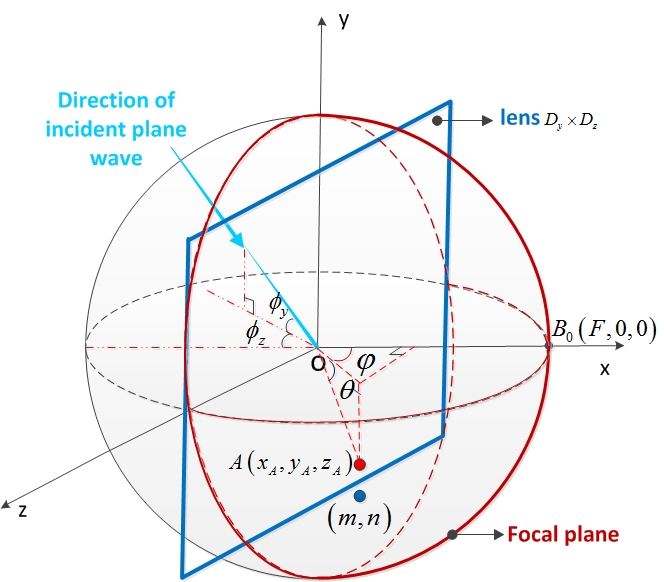}
\caption{Illustration of parameters concerning EM lens. The blue rectangle represents the EM lens with length $D_{y}$ and height $D_{z}$, and the red hemisphere surface represents the focal plane, with focal point $B_{0}$ and focal length $F$. The point on the focal plane represents an antenna with index $(m,n)$. The direction of the incident plane wave is determined by a 2-tuple angle $(\phi_{y},\phi_{z})$.}\label{sphere}
\end{figure}

The lens antenna array response is determined by its geometry. We consider
the EM lens as depicted in Fig. \ref{sphere}, where $D_{y}$ and $D_{z}$ are the length and height, respectively.
We assume that the thickness of the EM lens is negligible. The $M\times N$ antenna array is placed on the focal plane with radius $F$ shown as the red line in Fig. \ref{sphere}.
If $A(x_{A},y_{A},z_{A})$ is an arbitrary point on the focal plane, represented by the spherical coordinate system, then we obtain $x_{A}=F \cos\theta \cos\varphi$, $y_{A}=-F\sin\theta$, $z_{A}=-F\cos\theta \sin\varphi$  with $\theta,\varphi\in[-\frac{\pi}{2},\frac{\pi}{2}]$ shown in Figure \ref{sphere}.
The angles of the incident plane wave are described by 2-tuple $(\phi_{y},\phi_{z})$.
Let $\lambda$ represent the wave length of the incident plane wave, and the following new variables are introduced: $\tilde{D}_{y}=D_{y}/\lambda$, $\tilde{D}_{z}=D_{z}/\lambda$,
and $\sqrt{A}={\lambda}/{\sqrt{D_{y}D_{z}}}$.
Then, the receiver antenna array response is characterized by \cite{1}
\begin{equation} \label{n1}
 a_{\tilde{\theta}_{y},\tilde{\theta}_{z}}(\tilde{\phi}_{y},\tilde{\phi}_{z})
 =\sqrt{A} {\rm sinc}(\tilde{D}_{y}(\tilde{\theta}_{y}-\tilde{\phi}_{y})){\rm sinc}(\tilde{D}_{z}(\tilde{\theta}_{z}-\tilde{\phi}_{z})),
\end{equation}
where $\tilde{\theta}_{y}=\sin\theta$, $\tilde{\theta}_{z}=\sin\varphi\cos\theta$, $\tilde{\phi}_{y}=\sin\phi_{y}$ and $\tilde{\phi}_{z}=\sin\phi_{z}$.
Note that the antenna array response is the product of two ``sinc"  functions, and these functions
achieve their maximum values when $\tilde{\theta}_{y}=\tilde{\phi}_{y}$ and $\tilde{\theta}_{z}=\tilde{\phi}_{z}$, respectively. This property means that when we place an antenna near the point $(\tilde{\phi}_{y},\tilde{\phi}_{z})$ on the focal plane, the antenna receives the maximum power.

The focal plane has infinite points. However, in practice, we only place finite antennas on the focal plane.
We place $M\times N$ antennas on the focal plane, where the index of the antenna is $(m,n)$. Let $m =\tilde{\theta}_{y}\tilde{D}_{y}$ with $m\in\{0,\pm1,...,\pm\frac{M-1}{2}\}$ and $M=1+\lfloor2\tilde{D}_{y}\rfloor$, and $n=\tilde{\theta}_{z}\tilde{D}_{z}$ with $n\in\{0,\pm1,...,\pm\frac{N-1}{2}\}$ and $N=1+\lfloor2\tilde{D}_{z}\rfloor$.
As a result, the antenna array response sampled at the $(m,n)$-th antenna element
is given by
\begin{equation}\label{n}
a_{m,n}(\tilde{\phi}_{y},\tilde{\phi}_{z})=\sqrt{A} {\rm sinc}{\left(m-\tilde{D}_{y}\tilde{\phi}_{y}\right)}
{\rm sinc} {\left(n-\tilde{D}_{z}\tilde{\phi}_{z}\right)}.
\end{equation}
Clearly, when $m$ and $n$ are close to $\tilde{D}_{y}\tilde{\phi}_{y}$ and $\tilde{D}_{z}\tilde{\phi}_{z}$, respectively, $a_{m,n}(\tilde{\phi}_{y},\tilde{\phi}_{z})$ reaches the maximum value.  Therefore, we can observe that AoAs $(\tilde{\phi}_{y},\tilde{\phi}_{z})$ determines $a_{m,n}(\tilde{\phi}_{y},\tilde{\phi}_{z})$.
From (\ref{n}), we define the array response matrix
\begin{equation}\label{m6}
\mathbf{A}^{(l)}= \left[a_{m,n}\Big(\tilde{\phi}_{y}^{(l)},\tilde{\phi}_{z}^{(l)}\Big) \right]_{M\times N}.
\end{equation}
The number of antennas in the lens antenna array is $N_{a}=MN$.
Therefore, when (\ref{nn}) and (\ref{m6}) are combined, the channel matrix for the 3D lens system is given by
\begin{equation}\label{22}
\mathbf{H}=\sqrt{\frac{MN}{L+1}}\sum_{l=0}^{L}\alpha^{(l)}\mathbf{A}^{(l)}.
\end{equation}
If several paths with different AoAs exist, several large values should appear in the entries of $\mathbf{H}$.
By vectorizing $\mathbf{H}$, we obtain the beamspace channel vector $\mathbf{h}$ in (\ref{m1}). In Fig. \ref{sparsity}, we have shown the amplitude of $\mathbf{H}$ in which the channel possesses two paths.
The entries with stronger amplitude are painted with a darker color.
As expected, most entries can be ignored as zero because of the small proportion of the energy contribution.

The sparsity nature of $\mathbf{h}$ motivates us to use the compressive sensing technique for the channel estimation.
Given the selection network $\mathbf{W}$, the selected signal $\mathbf{r}\in\mathbb{C}^{Q\times 1}$ turns into
\begin{equation}\label{M2}
\mathbf{r}=\mathbf{W}\mathbf{y}.
\end{equation}
Substituting (\ref{m1}) into (\ref{M2}) results in
\begin{equation}\label{m3}
\mathbf{r}=\mathbf{W}\mathbf{h}s+\mathbf{n}^{\rm eff},
\end{equation}
where $\mathbf{n}^{\rm eff}=\mathbf{W}\mathbf{n}$ is the effective noise.
In this paper, $\mathbf{W}\in\mathbb{C}^{Q\times MN}$ is obtained by selecting a set of rows from a column- and row-permutated Hadamard matrix. Moreover, we assume $\mathbf{n}^{\rm eff} \sim {\cal N}(0, \Delta \mathbf{I}_{Q})$.
Since $s$ is known at the receiver side, we can easily remove its effect by multiplying $1/s$ on the right side of (\ref{m3}). Therefore, for ease of notation, we simply set $s = 1$ in this paper.
Then, we obtain
\begin{equation}\label{m5}
\mathbf{r} = \mathbf{W}\mathbf{h}+ \mathbf{n}^{\rm eff}.
\end{equation}
Our main task is to estimate $\mathbf{h}$ from $\mathbf{r}$ given the selection network matrix $\mathbf{W}$.

\section{SCAMPI-based Beamspace Channel Estimation}

In this section, we apply the SCAMPI method \cite{5} to the channel estimation.
Our motivation is based on the fact
that elements in the channel matrix are continuous like pixels on a 2D image; thus, changes between adjacent elements (i.e., horizontal and vertical neighbors) in the channel matrix should be subtle. Therefore, we can exploit
the correlation between the adjacent elements to improve channel estimation performance.

To apply SCAMPI to the channel estimation, we go through three key steps as summarized in Table \ref{Tab1}.
In \textbf{Step 1}, we formulate the estimation problem by introducing auxiliary variables so that the information on changes between adjacent entries in the channel matrix can be considered.
This additional information plays a critical role in improving the accuracy of the estimation results.
After formulating the estimation problem into a linear model, we can employ the Bayesian estimator.
In \textbf{Step 2}, we analyze the probability distributions of the channel responses and auxiliary variables because
the Bayesian estimator requires knowing the distribution of the concerned variables. Some of the prior parameters in \textbf{Step 2}, such as the noise variance, are unknown but required. In \textbf{Step 3}, we learn the noise variance through Bethe free energy, which improves the robustness of the SCAMPI algorithm to the model uncertainty. Detailed theoretical analysis of each step will be explained in the sequel.

\begin{table} 
\caption{Key steps of SCAMPI-based channel estimation method}\label{Tab1}
 \renewcommand\arraystretch{1.1}
\small
\begin{tabular}{cl}
\hline
\hline
\multicolumn{2}{l}{SCAMPI-based Channel Estimation Method}\\
\hline
\textbf{Step 1} & \textbf{Augmented system design:} Introduce a set of auxiliary
                  variables to construct augmented system model\\ &$\check{\mathbf{r}}\!\!=\!\!\check{\mathbf{W}}\check{\mathbf{h}}\!+\!\check{\mathbf n}^{\rm eff}$ shown in (\ref{ttt})
                  from (\ref{m5}).\\
\textbf{Step 2} & \textbf{Conditional mean and variance estimator deduction:}\\
                 & a) Analyze the priori probability distributions for the
                 elements in augmented channel vector $\check{\mathbf{h}}$ defined \\
                 & \hspace*{0.14in}in (\ref{s1}).\\
                 & b) Deduce the conditional mean estimator $f_{a_{i}}(\Sigma,R)$
                 and variance estimator $f_{v_{i}}(\Sigma,R)$ defined in (\ref{s17})\\
                 & \hspace*{0.14in}utilizing the corresponding probability distribution
                 functions.\\
\textbf{Step 3} & \textbf{Noise variance learning:} Learn the noise variance via
                  the Bethe free energy.\\
\hline
\end{tabular}
\end{table}

\subsection{System Model with Auxiliary Variables}

\begin{figure}
\centering
\includegraphics[scale=0.3,bb= 0 0 800 420]{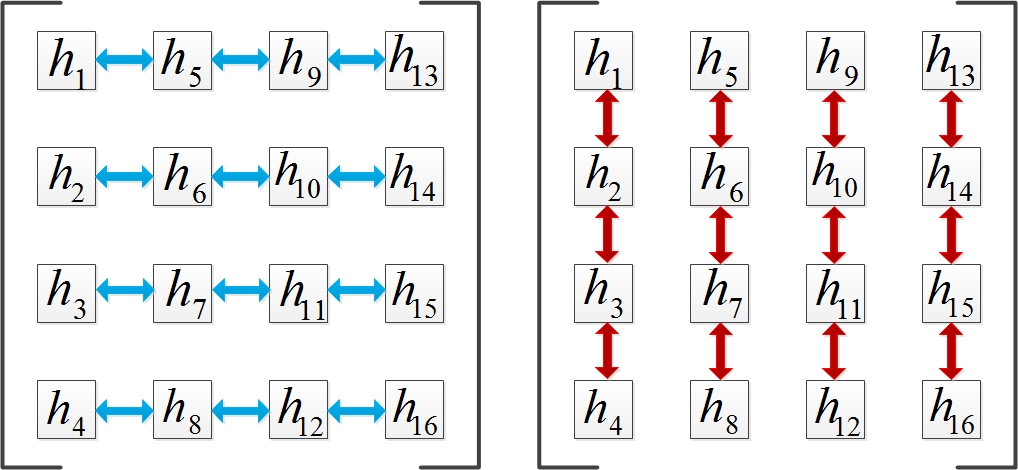}
\caption{Illustration of horizontal and vertical neighboring relations of elements in channel matrix $\mathbf{H}$, represented by blue and red arrows, respectively. In this figure, we assume $M=N=4$. The elements in $\mathbf{h}$ are placed in matrix $\mathbf{H}$ according to the order in this figure. }\label{gradient}
\end{figure}

To apply SCAMPI to the system model (\ref{m5}), we first introduce the auxiliary variables $\mathbf{d}$ that contain the information on changes between adjacent entries, i.e.,
\begin{equation}\label{s2}
\mathbf{d} \triangleq [h_{i}-h_{j}:(i,j)\in \mathbb{E}],
\end{equation}
where $\mathbb{E} $
is the set of all index pairs $(i,j)$ as $h_{i}$ and $h_{j}$ are adjacent.
To obtain an idea on $\mathbf{d}$, we take a ${4 \times 4}$ channel matrix ${\mathbf H}$ as an example illustrated in Fig. \ref{gradient}. Note that we have vectorized a ${4 \times 4}$ matrix $\mathbf{H}$ into a $16$-dimensional vector $\mathbf{h}$. The left and right parts of Fig. \ref{gradient} show horizontal and vertical neighboring relations of elements in the channel matrix $\mathbf{H}$, respectively.
Take a ${4 \times 4}$ channel matrix for example, let
\begin{equation}\label{dh}
{{\mathbf D}_{h}} =\begin{bmatrix}  1&0&0&0&-1&0&...&0 \\0&1&0&0&0&-1&...&0  \\\vdots&&&&&\vdots&\ddots&\vdots \\ 0&0&0&0&0&0&\cdots&-1 \end{bmatrix}
\end{equation}
be a $12\times16$ matrix. Then, ${\mathbf D}_{h} {\mathbf h}$ corresponds to $h_{i}-h_{j}$ in the left part of Fig. \ref{gradient}.
Similarly, let
\begin{equation}\label{dv}
 {\mathbf D}_{v} = \begin{bmatrix}  1&-1& 0&0&0&...&0&0 \\0&1&-1&0&0&...&0&0 \\\vdots&&&&\vdots&\ddots& &\vdots \\ 0&0&0&0&0&\cdots&1&-1 \end{bmatrix}
\end{equation}
be a $12\times16$ matrix.
Then, ${\mathbf D}_{v} {\mathbf h}$ corresponds to $h_{i}-h_{j}$ in the right part of Fig. \ref{gradient}.

We can infer from the above example that in general the row dimensions of ${\mathbf D}_{h}$ and ${\mathbf D}_{v}$ are ${M\times (N-1)}$ and ${N\times (M-1)}$, respectively. Therefore, we have
\begin{equation}\label{s5}
|\mathbb{E}|=M\times (N-1)+ N\times (M-1).
\end{equation}
Let
\begin{equation}\label{s4}
\mathbf{D}=\begin{bmatrix}{{\mathbf D}_{h}}\\{{\mathbf D}_{v}}\end{bmatrix} \in\mathbb{R}^{|\mathbb{E}|\times MN}.
\end{equation}
Then, we obtain the following relation:
\begin{equation} \label{Dh=d+e}
\mathbf{D} \mathbf{h} =\mathbf{d}.
\end{equation}
Combining (\ref{m5}) and (\ref{Dh=d+e}), we obtain
\begin{equation}\label{tt}
\begin{bmatrix} \mathbf{r} \\ \mathbf{0} \end{bmatrix}= \begin{bmatrix} \mathbf{W} & \mathbf{0}\\\mathbf{D} & -\mathbf{I} \end{bmatrix}\begin{bmatrix} \mathbf{h} \\ \mathbf{d} \end{bmatrix}+\begin{bmatrix}\mathbf{n}^{\rm eff}\\ \mathbf{e}\end{bmatrix},\\
\end{equation}
where $\mathbf{e} = {\mathbf 0}$.
Since the changes between adjacent channels should be subtle, we assume that $\mathbf{d}$ is a sparse vector, i.e., most of the elements of $\mathbf{d}$ are zero. To reflect the possibility error on this assumption, we consider $\mathbf{e}$ as an error vector. Specifically, $\mathbf{e}$ is assumed to be
a Gaussian vector with zero mean and variance $\Upsilon$, i.e., $\mathbf{e}\sim \mathcal{N}(\mathbf{0},\Upsilon{\mathbf I}_{|\mathbb{E}|})$. The setting of $\mathbf{e}$ can make the algorithm robust to the model error.

Let
\begin{equation}\label{s1}
\check{\mathbf{h}} \triangleq \begin{bmatrix}\mathbf{h}\\\mathbf{d}\end{bmatrix},
\quad \check{\mathbf{r}} \triangleq \begin{bmatrix}\mathbf{r}\\\mathbf{0}\end{bmatrix},
\quad \check{\mathbf n}^{\rm eff} \triangleq \begin{bmatrix}\mathbf{n}^{\rm eff}\\\mathbf{e}\end{bmatrix},
\quad \check{\mathbf{W}} \triangleq \begin{bmatrix}\mathbf{W}&\mathbf{0}\\\mathbf{D}&-\mathbf{I}\end{bmatrix}.
\end{equation}
The system model (\ref{tt}) can be expressed in compact form as
\begin{equation}\label{ttt}
\check{\mathbf{r}}=\check{\mathbf{W}}\check{\mathbf{h}}+\check{\mathbf n}^{\rm eff}.
\end{equation}

\vspace{-7mm}
\subsection{Conditional Mean and Variance Estimator}


After obtaining (\ref{ttt}), we can apply the (classical) AMP algorithm to obtain the Bayesian inference of $\check{\mathbf{h}}$ from $\check{\mathbf{r}}$.
To this end, one has to know the distribution of $\check{\mathbf{h}}$.
Vector $\check{\mathbf{h}}$ consists of two parts, $\mathbf{h}$ and $\mathbf{d}$, as shown in (\ref{s1}). According to the fact that the difference between adjacent elements is small,
we can assume a sparse non-informative parameter estimator (SNIPE) prior $P_{\rm SNIPE}$ for elements in $\mathbf{d}$ \cite{6}. The priori probability density function (pdf) of elements in $\mathbf{d}$ is given by
\begin{equation}\label{oo}
\begin{split}
P_{\rm SNIPE}(d_{i};\omega) \triangleq \displaystyle\lim_{\sigma\rightarrow\infty}\left(\frac{\rho(\sigma;\omega)}{\sigma}\mathcal{N}\left(\frac{d_{i}}{\sigma};0,\sigma^{2}\right)+(1-\rho(\sigma;\omega))\delta(d_{i})\right),\\
\end{split}
\end{equation}
where
\begin{equation}\label{s12}
\rho(\sigma;\omega)=\frac{\sigma}{\sigma+\mathcal{N}(0;0,\sigma^{2})\sqrt{2\pi\Sigma}e^{\omega}}
\end{equation}
is a proper scale and $\omega$ is a free parameter.
Therefore, the pdf of vector $\mathbf{d}$ is given by
\begin{equation}\label{s11}
P(\mathbf{d}) \triangleq \prod_{i=1}^{|\mathbb{E}|}P_{\rm SNIPE}(d_{i};\omega).
\end{equation}

We denote the pdf of elements in $\mathbf{h}$ as $P(h_{i})$, and the pdf of vector $\mathbf{h}$ is given by
\begin{equation}\label{s13}
P(\mathbf{h}) \triangleq \prod_{i=1}^{MN}P(h_{i}).
\end{equation}
Then, we discuss the specific probability distribution of $\mathbf{h}$.
As mentioned, the
elements in the channel matrix can be viewed as pixels for the image.
Based on the fact that the value of each pixel obeys uniform
distribution, we preliminarily assume that the elements of $\mathbf{h}$ follow a uniform distribution.
Therefore, the pdf of $\mathbf{h}$ is
\begin{equation}\label{u1}
P(\mathbf{h}) \triangleq \prod_{i=1}^{MN}\mathcal{U}(h_{i}),
\end{equation}
where $\mathcal{U}$ represents uniform distribution.

\begin{figure}
\centering
\includegraphics[scale=0.3,bb= 0 0 930 720]{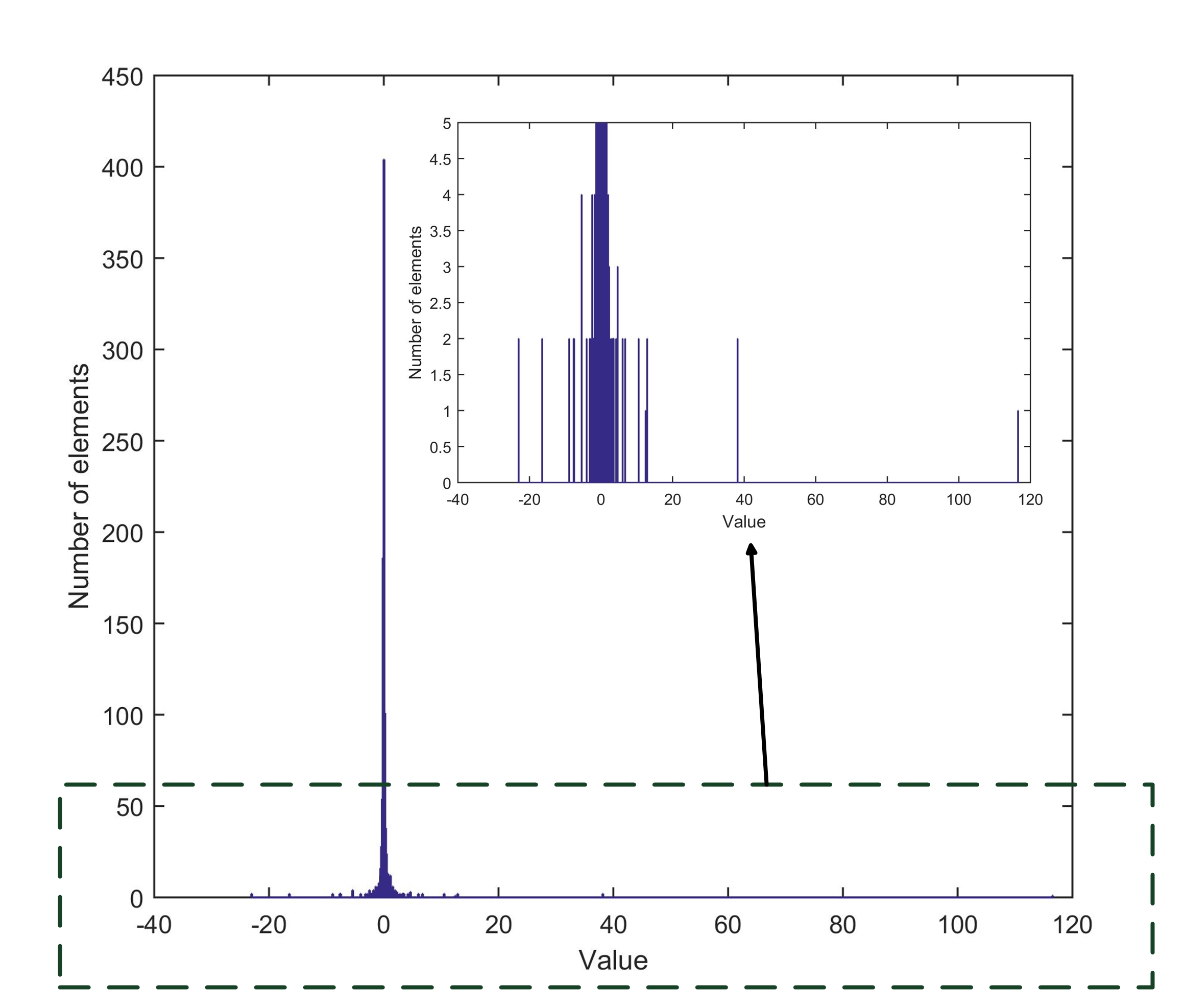}
\caption{Analysis of distribution for $\mathbf{h}$, where the vast majority of elements in $\mathbf{h}$ are nearly 0, and a tiny minority of elements are distributed in the range from -40 to 120.}\label{sparse-gaussian}
\end{figure}

In fact, as we observe in Fig. \ref{sparse-gaussian}, we can easily realize that the channel vector $\mathbf{h}$ is approximately sparse in which the vast majority of the elements in $\mathbf{h}$ are nearly zero.
Therefore, we further assume that the elements of $\mathbf{h}$ follow the sparse Gaussian distribution. The sparse Gaussian distribution of $\mathbf{h}$ is given by
\begin{equation}\label{g1}
P(\mathbf{h}) \triangleq \prod_{i=1}^{MN} \Big( \lambda\mathcal{N}(h_{i};a,v)+(1-\lambda)\delta(h_{i}) \Big),
\end{equation}
where $\lambda$ represents the sparsity rate. In contrast to the uniform distribution, the parameters $(\lambda,a,v)$ of the sparse Gaussian distribution in (\ref{g1}) are unknown.
Thus, we employ the EM algorithm embedded in SCAMPI to learn these parameters in the next section.
We will compare the performance difference caused by the uniform distribution and sparse Gaussian distribution.

The prior distribution function of $\check{\mathbf{h}}$ is given as follows
\begin{equation}\label{uu}
P(\check{\mathbf{h}})
=\prod_{i=1}^{MN}P(h_{i})\prod_{j=1}^{|\mathbb{E}|}P_{\rm SNIPE}(b_{j};\omega).
\end{equation}
To deduce the conditional mean and variance estimate of $\check{\mathbf{h}}$, we are required to know the posterior pdf $P(\check{\mathbf{h}}|\check{\mathbf{r}})$. Combining (\ref{s1})$-$(\ref{ttt}), and (\ref{uu}), we obtain
\begin{multline}\label{uuu}
P(\check{\mathbf{h}}|\check{\mathbf{r}}) \propto \prod_{\mu=1}^{Q} \mathcal{N}( h_{\mu};\mathbf{W}_{\mu}\mathbf{h},\Delta) \times \prod_{i=1}^{MN}P(h_{i})
 \times\prod_{\nu=1}^{|\mathbb{E}|} \mathcal{N}(d_{\nu};\mathbf{D}_{\nu}\mathbf{h},\Upsilon)P_{\rm SNIPE}(d_{\nu};\omega) ,
\end{multline}
where $\mathbf{W}_{\mu}$ and $\mathbf{D}_{\nu}$ represent $\mu$th row of $\mathbf{W}$ and $\nu$th row of $\mathbf{D}$, respectively.
Let $i$ represent the element index in vector $\check{\mathbf{h}}$. The Bayes-optimal way to estimate $\check{h}_i$ that minimizes the MSE is given by
\begin{equation}\label{hat_hi}
  \hat{\check{h}}_i = \int \check{h}_i P(\check{h}_i|\check{\mathbf{r}}) d \check{h}_i,
\end{equation}
where $P(\check{h}_i|\check{\mathbf{r}})$ denotes the marginal pdf of the $i$th variable under $P(\check{\mathbf{h}}|\check{\mathbf{r}})$. We observe that when $i> MN$, the priori probability distribution of elements in $\check{\mathbf{h}}$ follows SNIPE distribution. Meanwhile, when $i\leq MN$, the priori probability distribution of elements in $\check{\mathbf{h}}$ can be divided into two cases: Case (\romannumeral1) $\check{h}_{i}$ follows uniform distribution and Case (\romannumeral2) $\check{h}_{i}$ follows sparse Gaussian distribution.

The optimal Bayes estimation (\ref{hat_hi}) is not computationally tractable. To obtain an estimate of the marginal pdfs, we adopt the AMP algorithm in \cite{6}, which
is an iterative message passing algorithm. The implementation steps of AMP are listed in Algorithm 1. Lines 3--8 can be interpreted
as computing the mean and variance of $\check{\mathbf{h}}$ using the linear minimum mean-square error estimator under the measurement of $\check{\mathbf{r}}$.
Lines 9--10 can be interpreted as computing the posterior
mean and variance of $\check{\mathbf{h}}$ under mean $\{ R_{i} \}$ and variance $\{ \Sigma_{i} \}$. Specifically,
the updates of the local beliefs mean $a_{i}$ and variance $v_{i}$ are given by
\begin{equation}\label{f7}
a_{i}^{t+1}= f_{a_{i}}(\Sigma_{i}^{t+1},R_{i}^{t+1}),
\end{equation}
\begin{equation}\label{f8}
v_{i}^{t+1}= f_{v_{i}}(\Sigma_{i}^{t+1},R_{i}^{t+1}),
\end{equation}
where we introduce the conditional mean and variance estimators \cite{5} defined by
\begin{equation}\label{s17}
\begin{split}
&f_{a_{i}}(\Sigma,R) \triangleq E\{\check{h}_{i}|\Sigma,R \},\\
&f_{v_{i}}(\Sigma,R) \triangleq E\{\check{h}_{i}^{2}|\Sigma,R\}-f_{a_{i}}^{2}(\Sigma,R),\\
\end{split}
\end{equation}
where $R$ and $\Sigma$ are updated equivalent mean and variance. The expectation in (\ref{s17}) is with respect to
\begin{equation}
  \frac{ {\cal N}(\check{h}_{i}; R, \Sigma) P(\check{h}_i) }{ \int {\cal N}(\check{h}_{i}; R, \Sigma) P(\check{h}_i) d \check{h}_i}.
\end{equation}
The detailed derivations of (\ref{s17}) are listed in Appendix A.
The results are summarized as follows:
\begin{equation}\label{long1}
\begin{split}
f_{a_{i}}(\Sigma,R)=&\begin{cases}
\frac{R}{1+e^{\omega-\frac{R^{2}}{2\Sigma}}}, &\mbox{$i\!>\! MN$,}\\
R, &\mbox{$i\!\leq\! MN$ for Case (\romannumeral1),}\\
\frac{\frac{a}{v}+\frac{R}{\Sigma}}{(\frac{1}{v}+\frac{1}{\Sigma})(1+\eta)}, &\mbox{$i\!\leq\! MN$ for Case (\romannumeral2),}
\end{cases}\\
f_{v_{i}}(\Sigma,R)=&\begin{cases}
\frac{1}{1\!+\!e^{\omega-\frac{R^{2}}{2\Sigma}}}\!\!\left(\!\!\Sigma\!+\!\!\frac{R^{2}}{1\!+\!e^{\frac{R^{2}}{2\Sigma}-\omega}}\!\!\right), &\mbox{$\!i>\! MN$,}\\
\Sigma , &\mbox{$i\!\!\leq\!\! MN$ for Case (\romannumeral1),}\\
\frac{\frac{1}{\frac{1}{v}+\frac{1}{\Sigma}}+\Big(\frac{\frac{a}{v}+\frac{R}{\Sigma}}{\frac{1}{v}+\frac{1}{\Sigma}}\Big)^2\frac{\eta}{1+\eta}}{1+\eta}, &\mbox{$i\!\!\leq\!\! MN$ for Case (\romannumeral2),}
\end{cases}
\end{split}
\end{equation}
where
\begin{equation}\label{s21}
\eta \triangleq \frac{1-\lambda}{\lambda}\sqrt{\frac{v+\Sigma}{\Sigma}}e^{\frac{(a-R)^2}{2(v+\Sigma)}-\frac{R^{2}}{2\Sigma}}.
\end{equation}

\subsection{Noise Variance Learning}
Although AMP is a powerful method, it does not always converge to a solution. According to \cite{bethe}, the convergence properties of the AMP algorithm can be increased by estimating the variances of the noise and sparsity error.
In our proposed algorithm, $\Delta$ and $\Upsilon$ can be estimated by using the Bethe free energy. Let
\begin{equation}\label{s14}
\check{\Delta}_{\mu}=\begin{cases}
\Delta &\mbox{for $\mu=1,2\ldots, Q$,}\\
\Upsilon &\mbox{for $\mu= Q+1, \ldots Q+|\mathbb{E}|$.}
\end{cases}
\end{equation}
\cite{5} provides the function of Bethe free energy, and $\check{\Delta}_{\mu}$ is chosen to minimize the Bethe free energy. Following \cite{5}, we obtain
\begin{equation}\label{www}
\check{\Delta}^{2}_{\mu}-\delta_{\mu}^{2}\check{\Delta}_{\mu}-\delta_{\mu}^{2}\sum_{i=1}^{MN+|\mathbb{E}|}\check{W}^{2}_{\mu,i}v_{i}=0,
\end{equation}
where we define
\begin{equation}\label{s15}
\delta_{\mu}=\check{r}_{\mu}-\sum_{i=1}^{MN+|\mathbb{E}|}\check{W}_{\mu,i}a_{i}.
\end{equation}
The feasible solution of (\ref{www}) is given by
\begin{equation}\label{s16}
\check{\Delta}_{\mu}= 0.5\left(\delta_{\mu}^{2}+\delta_{\mu}\sqrt{\delta_{\mu}^{2}+4\sum_{i=1}^{MN+|\mathbb{E}|}\check{W}_{\mu,i}^{2}v_{i}}\right).
\end{equation}
Thus, the iterative process for $\check{\Delta}_{\mu}$ is given by
\begin{equation}\label{delta}
\check{\Delta}_{\mu}^{t+1}=0.5\left((\delta_{\mu}^{t})^2+\delta_{\mu}^{t}\sqrt{(\delta_{\mu}^{t})^2+4\sum_{i=1}^{MN+|\mathbb{E}|} \check{W}_{\mu,i}^{2} v_{i}^{t}}\right).
\end{equation}
Lines 12 and 13 of the algorithm 1 can be interpreted as computing the equivalent noise variance $\check{\Delta}_{\mu}$. The estimation of $\check{\Delta}_{\mu}$ makes SCAMPI more robust than other methods to uncertainties on the channel.

\begin{algorithm}[t]\label{alg1}
\caption{SCAMPI Algorithm} 
\hspace*{0.02in} {\bf Input:} 
$\check{\mathbf{r}}$ and $\check{\mathbf{W}}$\\
\hspace*{0.02in} {\bf Output:} 
$\check{\mathbf{h}}_{\rm est}$
\begin{algorithmic}[1]
\raggedright
\State $t=0$;\ $\tau=1+\epsilon$ 
\While{$t\leq t_{\max}\ and\  \tau>1+\epsilon$} 
\State $ \tilde{\Theta}_{\mu}^{t+1}=\Sigma^{MN+|\mathbb{E}|}_{i=1}\check{W}_{\mu,i}^{2}v_{i}^{t}$
\State $ \tilde{\Phi}_{\mu}^{t+1}=\Sigma^{MN+|\mathbb{E}|}_{i=1}\check{W}_{\mu,i}a_{i}^{t}-\tilde{\Theta}_{\mu}^{t+1}\frac{\check{r}_{\mu}-\Phi_{\mu}^{t}}{\check{\Delta}_{\mu}^{t}+\Theta_{\mu}^{t}} $
\State $\Phi_{\mu}^{t+1}=\beta\Phi_{\mu}^{t}+(1-\beta)\tilde{\Phi}_{\mu}^{t+1}$
\State $ \Theta_{\mu}^{t+1} = \beta \Theta_{\mu}^{t}+(1-\beta)\tilde{\Theta}_{\mu}^{t+1}$
\State $ R_{i}^{t+1}= \frac{\Sigma^{Q+|\mathbb{E}|}_{\mu=1}\check{W}_{\mu,i}\frac{\check{r}_{\mu}-\Phi_{\mu}^{t+1}}{\check{\Delta}_{\mu}^{t}+\Theta_{\mu}^{t+1}}}{\Sigma^{Q+|\mathbb{E}|}_{\mu=1}\frac{\check{W}_{\mu,i}^{2}}{\check{\Delta}_{\mu}^{t}+\Theta_{\mu}^{t+1}}}+a_{i}^{t}$
\State $  \Sigma_{i}^{t+1}=\frac{1}{\Sigma^{Q+|\mathbb{E}|}_{\mu=1}\frac{\check{W}_{\mu,i}^{2}}{\check{\Delta}_{\mu}^{t}+\Theta_{\mu}^{t+1}}}$
\State $  a_{i}^{t+1}= f_{a_{i}}(\Sigma_{i}^{t+1},R_{i}^{t+1})$
\State $  v_{i}^{t+1}= f_{v_{i}}(\Sigma_{i}^{t+1},R_{i}^{t+1})$
\State $  \delta_{\mu}^{t+1} = \check{r}_{\mu}-\Sigma^{MN+|\mathbb{E}|}_{i=1}\check{W}_{\mu,i}a_{i}^{t+1}$
\State $  \tilde{\Delta}_{\mu}^{t+1} = 0.5(\delta_{\mu}^{t+1})^{2}$
\hspace*{0.2in}$+0.5\delta_{\mu}^{t+1}\sqrt{(\delta_{\mu}^{t+1})^{2}+4\Sigma^{MN+|\mathbb{E}|}_{i=1}\check{W}_{\mu,i}^{2}v_{i}^{t+1}}$
\State $  \check{\Delta}_{\mu}^{t+1} =\alpha\check{\Delta}_{\mu}^{t}+(1-\alpha)\tilde{\Delta}_{\mu}^{t+1}$
\State $  \tau=\frac{1}{MN}\|\mathbf{a}^{t+1}-\mathbf{a}^{t}\|^{2}_{2}$
\State $ t =t+1$
\EndWhile
\State \Return $\check{\mathbf{h}}_{\rm est}=\mathbf{a}^{t}$
\end{algorithmic}
\end{algorithm}

Having explained the steps in Table \ref{Tab1}, we illustrate the SCAMPI algorithm, which is shown in Algorithm 1, where $\mu\in\{1,2,...,Q\}$ and $i\in\{1,2,...,MN+|\mathbb{E}|\}$; $\alpha$ and $\beta$ are damping factors $(0\leq\alpha,\beta<1)$.
The reasonable initialization is given in Table \ref{Tab2},
where $a_{i}^{0}$, $v_{i}^{0}$, and $\check{\Delta}_{\mu}^{0}$ are the initial of the posterior mean, the posterior variance, and the noise variance. Note that the initial setting is used only in the beginning. With the increase in the number of iterations, the influence caused by the initial values decreases.
After $t_{\max}$ times iteration or when $\tau=\frac{1}{MN}\|\mathbf{a}^{t+1}-\mathbf{a}^{t}\|^{2}_{2}$ is smaller than the given threshold, the algorithm stops. The posterior mean $\mathbf{a}^{t}$ is regarded as the estimation of vector $\check{\mathbf{h}}$, which is denoted as $\check{\mathbf{h}}_{\rm est}=[\mathbf{h}_{\rm est}^T \ , \mathbf{d}_{\rm est}^T ]^{T}$. Then, the estimated channel vector is $\mathbf{h}_{\rm est}$.
\begin{table} [!hbp]
\caption{Reasonable initialization for the SCAMPI algorithm}\label{Tab2}
\center
\renewcommand\arraystretch{2}
\begin{tabular}{l|c|c|c|c|c}
\hline
\hline
Variable name& $a_{i}^{0}$ & $v_{i}^{0}$ & $\Theta_{\mu}^{0}$ & $W_{\mu}^{0}$ & $\check{\Delta}_{\mu}^{0}$\\
\hline
Initial value& $0$ & $0.1$ & $\frac{MN}{10\times Q}$ & $0$ & $0.1$ \\
\hline
\end{tabular}
\end{table}

The SCAMPI algorithm assumes that the prior distribution of the channel is known. We apply the EM learning algorithm in the next section to learn the varied prior distribution of the channel at each iteration. 

\section{EM Learning}
In this section, we apply the EM algorithm \cite{6,em3,em1,em2} to learn the parameters $(\lambda,a,v)$ in sparse Gaussian pdf $P(\mathbf{h})=\prod_{i=1}^{MN}(\lambda\mathcal{N}(h_{i};a,v)+(1-\lambda)\delta(h_{i}))$. To simplify the expression, we denote a set of parameters as $\mathbf{q} \triangleq [\lambda,a,v]$. At the beginning of each iteration, we update parameter $\mathbf{q}$ by using the EM learning algorithm to further improve the performance of the SCAMPI method.

\subsection{Retrospect of EM Learning Algorithm}
The EM algorithm is an iterative method in statistics to increase a lower bound on the likelihood $P(\mathbf{r};\mathbf{q})$, and then find (locally) the maximum likelihood estimates of parameters $\mathbf{q}$. In the given system model, $\mathbf{r}$ is a set of observed data, and $\mathbf{h}$ is a set of unobserved data to be estimated. For arbitrary pdf $\hat{P}(\mathbf{h})$, the EM algorithm is demonstrated as follows:
\begin{equation}
\begin{split}
\ln P(\mathbf{r};\mathbf{q})\!=\!\int \hat{P}(\mathbf{h})\ln P(\mathbf{r};\mathbf{q}) d {\mathbf{h}} \!=\!\underbrace{E_{\hat{P}(\mathbf{h})}\{\ln P(\mathbf{h},\mathbf{r};\mathbf{q})\}+H(\hat{P})}_{\mbox{lower bound}} \!+\!D(\hat{P}\parallel P_{\mathbf{h}|\mathbf{r}}(\cdot|\mathbf{r};\mathbf{q})),
\end{split}
\end{equation}
where $E_{\hat{P}(\mathbf{h})}\{\cdot\}$ represents expectation over $\mathbf{h}\sim\hat{P}(\mathbf{h})$, $H(\hat{P})$ represents the entropy of pdf $\hat{P}(\mathbf{h})$, and $D(\hat{P}\parallel P_{\mathbf{h}|\mathbf{r}}(\cdot|\mathbf{r};\mathbf{q}))$ represents the KL divergence between pdf $\hat{P}(\mathbf{h})$ and $P_{\mathbf{h}|\mathbf{r}}(\mathbf{h}|\mathbf{r};\mathbf{q})$, which is nonnegative.
The EM algorithm seeks to find the (locally) maximum likelihood estimates of parameters by iteratively using two steps: (E step) calculates the expected pdf of $\hat{P}(\mathbf{h})$ to maximize the lower bound, and (M step) finds the parameters $\mathbf{q}$ that maximize the lower bound for fixed $\hat{P}(\mathbf{h})$. For the E step, if $\hat{P}(\mathbf{h})=P_{\mathbf{h}|\mathbf{r}}(\mathbf{h}|\mathbf{r};\mathbf{q})$, we obtain $D(\hat{P}\parallel P_{\mathbf{h}|\mathbf{r}}(\cdot|\mathbf{r};\mathbf{q})) =0$; thus, the lower bound reaches the maximum value. In addition, for the M step, the expected $\mathbf{q}$, which maximizes the lower bound, would clearly be $\mathbf{q}=\arg\max \hat{E}\{\ln P(\mathbf{h},\mathbf{r};\mathbf{q})|\mathbf{r};\mathbf{q}\}$ for fixed $\hat{P}(\mathbf{h})=P_{\mathbf{h}|\mathbf{r}}(\mathbf{h}|\mathbf{r};\mathbf{q})$, where $\hat{E}\{\cdot\}$ denotes expectation over $\mathbf{h}\sim P_{\mathbf{h}|\mathbf{r}}(\mathbf{h}|\mathbf{r};\mathbf{q})$. Owing to the difficulty of joint optimization, one parameter in set $\mathbf{q}$ is calculated at a time while the others are fixed.

\subsection{Updates of Sparsity Rate $\lambda$, Mean $a$, and Variance $v$}
The set of parameters $\mathbf{q}$ can be updated by applying the EM algorithm. Considering the previous analysis, we use
\begin{equation}\label{e1}
\mathbf{q}=\arg\max \hat{E}\{\ln P(\mathbf{h},\mathbf{r};\mathbf{q})|\mathbf{r};\mathbf{q}\}.
\end{equation}
We update one argument of $\mathbf{q}$ each time by fixing the others.
(See Appendix B for the derivations.)
The closed-form expression for the EM update of $\lambda$, $a$, and $v$ are given by
\begin{equation}\label{e5}
\lambda^{t+1}=\frac{1}{MN}\sum_{i=1}^{MN}\pi_{i},
\end{equation}
\begin{equation}\label{e6}
a^{t+1}=\frac{1}{MN\lambda^{t+1}}\sum_{i=1}^{MN}\pi_{i}\gamma_{i},
\end{equation}
and
\begin{equation}\label{e7}
v^{t+1}=\frac{1}{MN\lambda^{t+1}}\sum_{i=1}^{MN}\pi_{i}(\nu_{i}+(\gamma_{i}-a^{t+1})^{2}),
\end{equation}
respectively, where $\pi_{i}$, $\gamma_{i}$, and $\nu_{i}$  are $P_{h_{i}|\mathbf{r}}(h_{i}|\mathbf{r};\mathbf{q}^{t})$-dependent quantities defined in Appendix B.

By embedding the derived EM updates into the SCAMPI algorithm in Section \uppercase\expandafter{\romannumeral3}, we are able to develop a more robust EM learning-based SCAMPI algorithm, which is summarized in Algorithm 2. Lines 3-5 in the algorithm are EM updates of the unknown parameters. One would simply run the EM learning-based SCAMPI algorithm with initializations illustrated in Section \uppercase\expandafter{\romannumeral3}.

\begin{algorithm}[t]\label{alg2}
\caption{EM Learning-based SCAMPI Algorithm} 
\hspace*{0.02in} {\bf Input:} 
$\check{\mathbf{r}}$ and $\check{\mathbf{W}}$\\
\hspace*{0.02in} {\bf Output:} 
$\check{\mathbf{h}}_{\rm est}$
\begin{algorithmic}[1]
\raggedright
\State $t=0$;\ $\tau=1+\epsilon$ 
\While{$t\leq t_{\max}\ and\  \tau>1+\epsilon$} 
\State $\lambda_{j\in[1,MN]}^{t+1}=\frac{1}{MN}\sum_{i=1}^{MN}\pi_{i}$
\State $a_{j\in[1,MN]}^{t+1}=\frac{1}{MN\lambda_{j}^{t+1}}\sum_{i=1}^{MN}\pi_{i}\gamma_{i}$
\State $v_{j\in[1,MN]}^{t+1}=\frac{1}{MN\lambda_{j}^{t+1}}\sum_{i=1}^{MN}\!\!\pi_{i}(\nu_{i}\!+\!(\gamma_{i}-a_{i}^{t+1})^{2})$

\State Compute step 3$-$13 in \textbf{Algorithm 1}
\State $ t =t+1$
\EndWhile
\State \Return $\check{\mathbf{h}}_{\rm est}=\mathbf{a}^{t}$
\end{algorithmic}
\end{algorithm}

\section{Numerical Results}
In this section, we conduct simulations to investigate the performance of the SCAMPI algorithm. The fundamental parameters in all the simulations are the same, the maximum iteration $t_{\max}$ is set to $300$, and the threshold $\tau$ is set to $10^{-20}$. The number of path $L$ is set to 3, equivalent lens length $\tilde{D}_{y}=\tilde{D}_{z}=12$, and wavelength $\lambda=1$.
We use the normalized mean-squared error (NMSE)
\begin{equation}\label{v1}
{\rm NMSE} = E  \left \{   \frac{{\|\mathbf{h}_{\rm est}-\mathbf{h}\|_{2}}^{2}}{{\|\mathbf{h}\|_{2}}^{2}} \right \}
\end{equation}
to evaluate the performance of each algorithm.

\subsection{SCAMPI with Uniform Distribution}
In this subsection, we compare the performance of the SCAMPI algorithm under uniform distribution with SD algorithm \cite{3} at the same simulation conditions. The SD algorithm estimates the main values of the channel, shown as the squares in Fig. \ref{sparsity}, and views the entries outside squares as 0. First, in the SD algorithm, the strongest path (LoS path) of the $\mathbf{h}$ is estimated. Second, the index of valid entries corresponding to the strongest path is recorded. Third, the influence of the strongest path is subtracted and the next strongest path (NLoS path) is estimated by the same method. Through this analogy, the collection of all of the valid entry positions and corresponding measurement matrix $\mathbf{\tilde{W}}$ are obtained. Finally, the LS algorithm is applied to estimate the channel as follows:
\begin{equation}\label{v}
\tilde{\mathbf{h}}_{\rm est}=(\tilde{\mathbf{W}}^{H}\tilde{\mathbf{W}})^{-1}\tilde{\mathbf{W}}\mathbf{r}.
\end{equation}

\begin{figure}[htbp!]
\centering
\includegraphics[scale=0.19,bb=0 0 1400 1284]{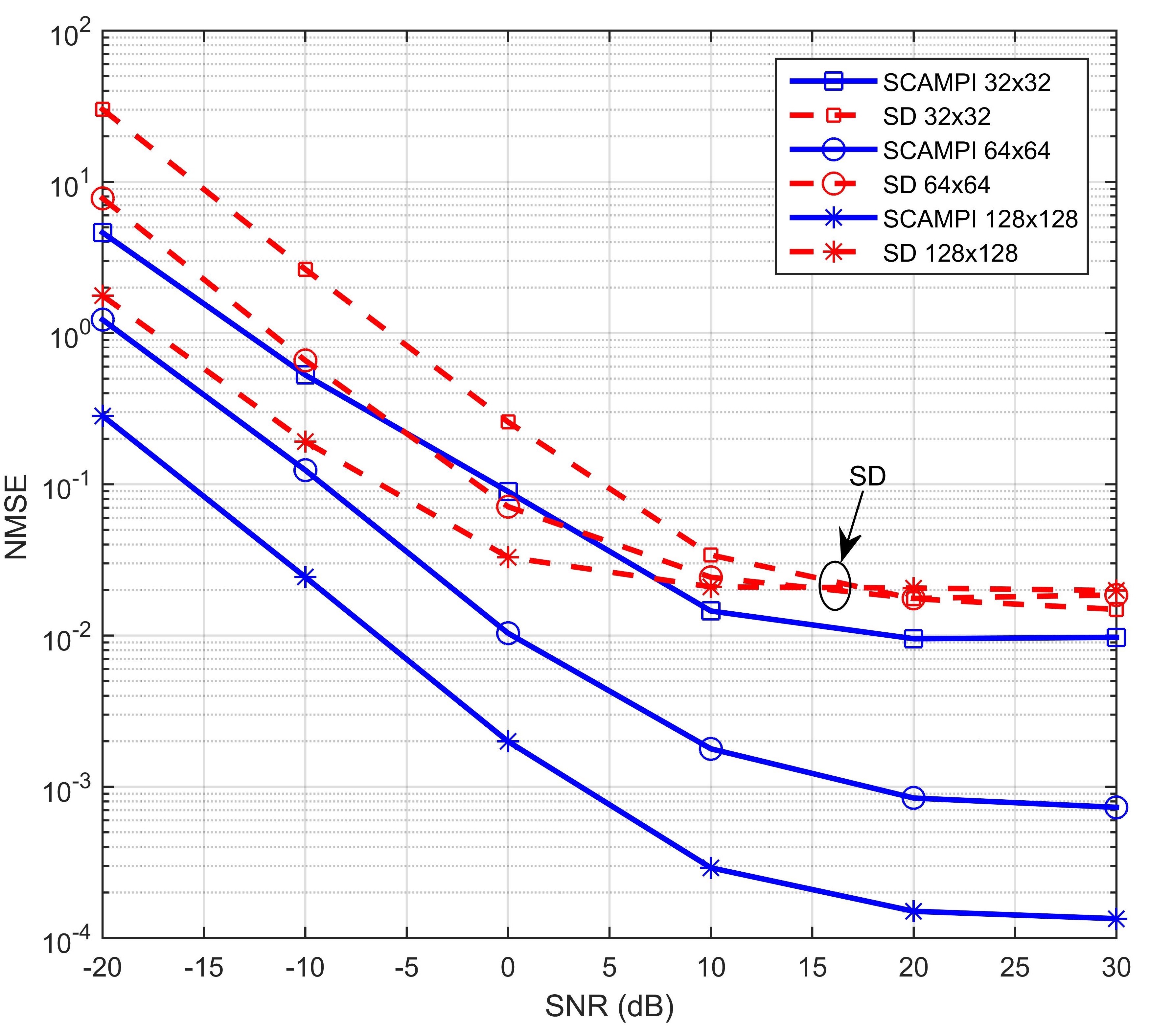}
\caption{Comparison of NMSE performance curves for SCAMPI algorithm under uniform distribution and SD algorithm. Results are shown for lens antenna array sizes of ${32 \times 32}$, ${64 \times 64}$, and ${128 \times 128}$. Correlation parameters are $\tau=10^{-20}$, $t_{\max}=300$, $L=3$,$\tilde{D}_{y}=\tilde{D}_{z}=12$, and $\lambda=1$; SD square size is $8\times8$. }\label{f1}
\end{figure}

Fig. \ref{f1} plots the NMSE according to (\ref{v1}) of the SCAMPI algorithm and SD algorithm at SNR from $-20$dB to $30$dB. By comparing the performance of the SCAMPI algorithm with the SD algorithm at the same size of the lens antenna array, we find out that the SCAMPI algorithm for channel estimation is much better than the SD algorithm for several orders especially at high SNR.
When we increase the size of the lens antenna array from $32\times 32$ to $64\times 64$ and $128\times 128$, the performance of the SCAMPI algorithm becomes better stably, with NMSE converging from around $10^{-2}$ to $10^{-3}$ and $10^{-4}$, respectively.
Although the performance of the SD algorithm improves by increasing the array size at low SNR, NMSE converges at the same value around $2\times10^{-2}$ when SNR exceeds $20$dB.
The result is reasonable, because the SD algorithm only utilizes the sparsity feature of $\mathbf{h}$, and the clustering feature of the path is not considered in the SD algorithm. The SCAMPI considers the clustering feature of the path. The existence of the augmented part $\mathbf{d}$ in (\ref{tt}) in the SCAMPI method enhances the accuracy of the estimation because the vector $\mathbf{d}$ includes information on the value difference between the adjacent entries in the channel matrix.

\subsection{SCAMPI with Sparse Gaussian Distribution}
In the previous subsection, SCAMPI is conducted under uniform distribution. To further improve the SCAMPI algorithm performance, we substitute sparse Gaussian distribution for uniform distribution. Furthermore, we embed EM learning in the SCAMPI algorithm to learn the parameters in pdf of sparse Gaussian distribution. For convenience, the SCAMPI algorithm conducted under uniform distribution is called Uniform-SCAMPI and the SCAMPI algorithm conducted under sparse Gaussian distribution with EM learning is called EM-Gaussian-SCAMPI. The simulation parameters are the same as those used in the previous experiment.

\begin{figure}[htbp!]
\centering
\includegraphics[scale=0.19,bb=0 0 1400 1284]{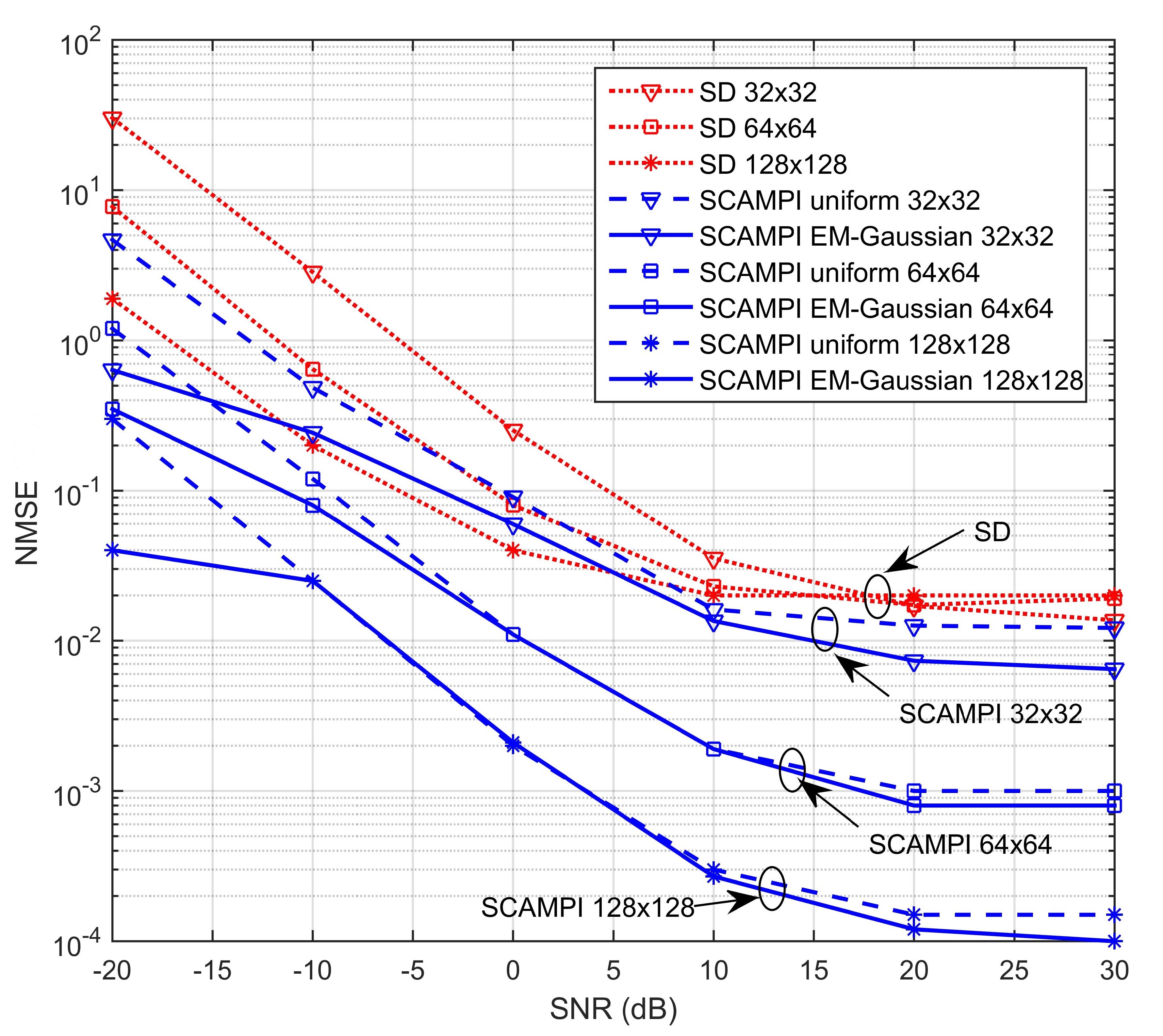}
\caption{Comparison of NMSE performance curves for EM-Gaussian-SCAMPI and Uniform-SCAMPI. Results are shown for lens antenna array sizes of ${32 \times 32}$, ${64 \times 64}$, and ${128 \times 128}$. Correlation parameters are $\tau=10^{-20}$, $t_{\max}=300$, $L=3$, $\tilde{D}_{y}=\tilde{D}_{z}=12$ and $\lambda=1$.}\label{f2}
\end{figure}

Fig. \ref{f2} shows the comparisons between EM-Gaussian-SCAMPI and Uniform-SCAMPI made for different antenna array sizes of ${32\times 32}$, ${64\times 64}$, and ${128\times 128}$, respectively.
The NMSE performance of EM-Gaussian-SCAMPI is much better than that of Uniform-SCAMPI when the size of the antenna array is $32\times32$. When SNR is equal to $-20$dB, the NMSE of EM-Gaussian-SCAMPI is $5\times10^{-1}$, whereas the NMSE of Uniform-SCAMPI is $5\times10^{0}$. Moreover, the NMSE of EM-Gaussian-SCAMPI converges at approximately $5\times10^{-3}$, while the NMSE of Uniform-SCAMPI converges at around $10^{-2}$. The EM-Gaussian-SCAMPI is much better than Uniform-SCAMPI at low SNR although the performance gap between EM-Gaussian-SCAMPI and Uniform-SCAMPI is gradually narrowed with the increase of the antenna array size at high SNR.

Before ending this subsection, we provide further discussions of the preceding simulation results. Sparse Gaussian distribution is closer to the real priori probability distribution of channel responses than uniform distribution, as we have analyzed in the previous section. We do not require complete knowledge of priori probability distribution parameters for channel responses, which are learned and updated as part of the estimation procedure. Therefore, the discussions support the argument on the performance of SCAMPI under sparse Gaussian distribution with EM learning. The performance of SCAMPI reaches the limit along with the increase of the antenna array size, which explains the closeness of performance between EM-Gaussian-SCAMPI and Uniform-SCAMPI at high SNR with a large antenna array size.

\vspace{-0.2em}
\subsection{Phase Shifter Reduction}
\begin{figure}[htbp!]
\centering
\includegraphics[scale=0.19,bb=0 0 1400 1284]{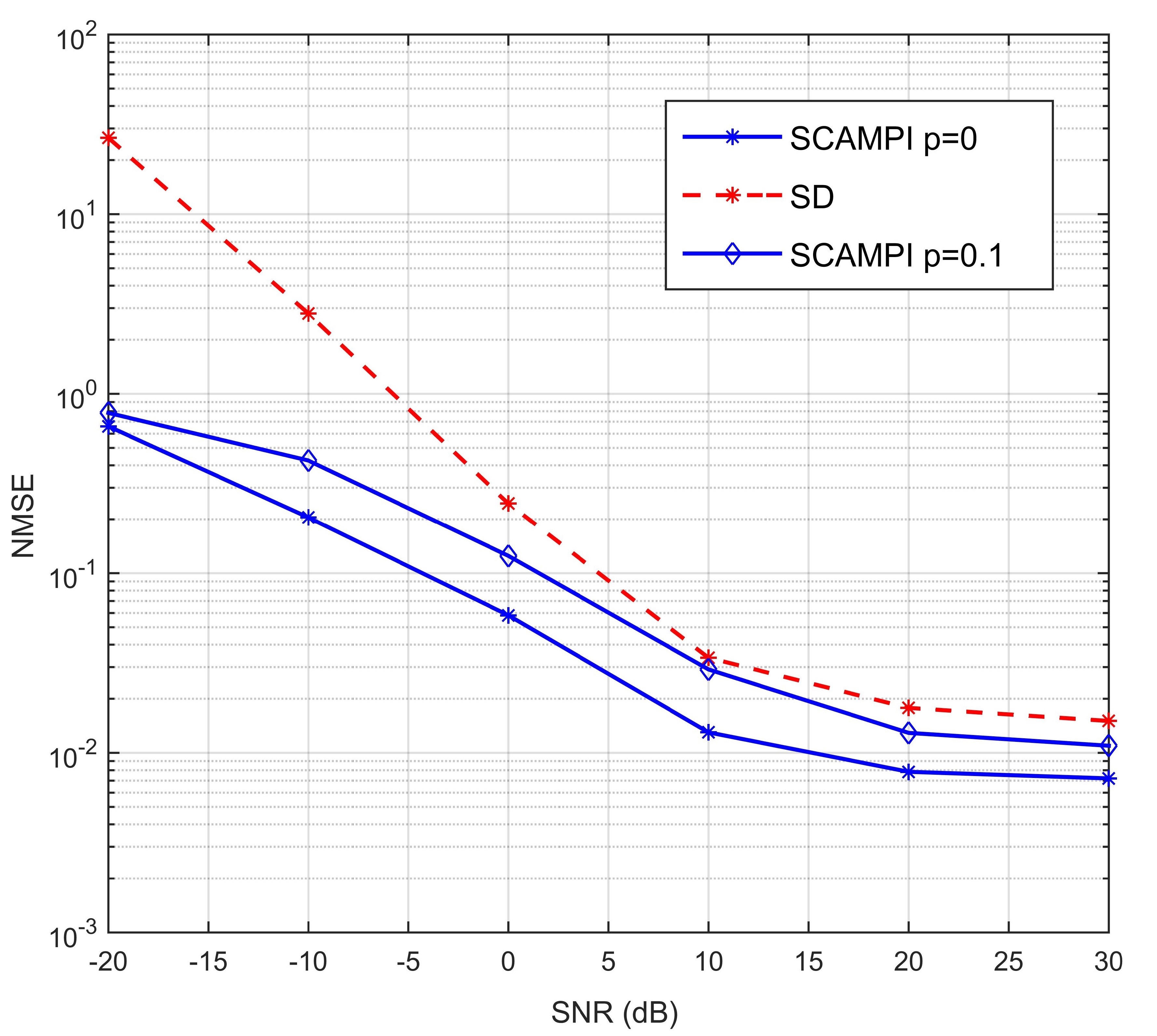}
\caption{Comparison of NMSE performance curves for EM-Gaussian-SCAMPI with phase-shifter-reduced adaptive selection network and fully connected phase-shifter selection network. Results are shown for lens antenna array size of ${32 \times 32}$. Correlation parameters are $\tau=10^{-20}$, $t_{\max}=300$, $L=3$, $\tilde{D}_{y}=\tilde{D}_{z}=12$, $\lambda=1$ and ratio $p=0.1$.}\label{ps2}
\end{figure}

In this section, we study the effect of the phase-shifter-reduced adaptive selection network structure on the performance of EM-Gaussian-SCAMPI.  The measurement matrix $\mathbf{W}$ is a selection network formed by phase shifters. The phase shifter is a device that consumes a large amount of power. To reduce the power consumption, we propose a new phase-shifter-reduced measurement matrix structure, as shown in Fig. \ref{ps:b}, in which a random part of the phase shifters are disconnected from the entire network. Let $p$ denote a ratio of disconnected phase shifters in total phase shifters. The simulations are designed to study the performance of the EM-Gaussian-SCAMPI for the phase-shifter-reduced adaptive selection network structure.
Fig. \ref{ps2} shows that the NMSE increases by approximately $4\times10^{-3}$ at SNR equal to $30$dB with antenna array size of $32\times32$ when the number of phase shifters decreases by $10\%$. Compared with a $10\%$ reduction in power consumption of the selection network, the performance degradation is negligible. 

\section{Conclusion}
The introduction of the 3D lens antenna array architecture in mmWave communication system brought a new concept (or structure) to the channel estimates. We utilized the focusing property of the lens antenna array and the sparsity of the corresponding channel response matrix. These properties showed that the channel estimation problem can be solved by using cosparse technique in compressive image reconstruction. In particular, we applied the SCAMPI algorithm originally used for image reconstruction to the channel estimation problems. The results showed that SCAMPI method can significantly outperform the existing SD method for several orders in accuracy. In particular, with the increase in the size of the antenna array, the performance of the SCAMPI algorithm obviously improves.

In image signal process problems, the SCAMPI algorithm uses uniform distribution as priori probability distribution. According to our analysis, the channel responses are closer to sparse Gaussian distribution than uniform distribution. Thus, we substituted sparse Gaussian distribution for uniform distribution as priori probability distribution of channel responses, and embed the EM learning algorithm into the SCAMPI algorithm to obtain precise values of prior parameters. Simulation results revealed that the EM-Gaussian-SCAMPI significantly outperforms the Uniform-SCAMPI especially for smaller antenna array size.

Finally, we found out that the power consumption for phase shifters under conventional selection network is large. Therefore, we proposed a new phase-shifter-reduced architecture for the selection network. The proposed EM-Gaussian-SCAMPI channel estimation algorithm was also used to study the effect of reducing phase shifters. Results illustrated that the number of phase shifters can be reduced by  $10\%$
without evident degradation in performance. The preceding discussions have shown that the EM-Gaussian-SCAMPI algorithm is a robust channel estimation method worth studying.

\ifCLASSOPTIONcaptionsoff
  \newpage
\fi

\begin{appendices}
\section{}
In this Appendix, we derive the conditional mean and variance estimator by conducting a separate treatment for the three cases: 1) $i> MN$, 2) $i\leq MN$ with $\mathbf{h}$ following uniform distribution, and 3) $i\leq MN$ with $\mathbf{h}$ following sparse Gaussian distribution.

\vspace{0.7em}
\emph{Case} \uppercase\expandafter{\romannumeral1}: In this case, $i> MN$ and $\mathbf{d}$ follow the SNIPE distribution.
According to the definition of conditional mean and variance estimator in (\ref{s17}),
 the SNIPE-specified estimator can be rewritten as follows:
\begin{equation}\label{ap1}
f_{a_{i}}(\Sigma,R)
=\frac{\int d_{i}P_{\rm SNIPE}(d_{i};\omega)\mathcal{N}(d_{i};R,\Sigma)dd_{i}}{\int P_{\rm SNIPE}(d_{i};\omega)\mathcal{N}(d_{i};R,\Sigma)dd_{i}},
\end{equation}
and
\begin{equation}\label{ap2}
f_{v_{i}}(\Sigma,R)
\!=\!\frac{\!\int\! d^{2}_{i}P_{\rm SNIPE}(d_{i};\omega)\mathcal{N}(d_{i};R,\Sigma)dd_{i}}{\!\int\! P_{\rm SNIPE}(d_{i};\omega)\mathcal{N}(d_{i};R,\Sigma)dd_{i}}-f_{a_{i}}^{2}(\Sigma,R).\\
\end{equation}
Substituting (\ref{oo}) into (\ref{ap1}) and (\ref{ap2}) yields
\begin{equation}\label{ap3}
\begin{split}
f_{a_{i}}(\Sigma,R)
\!=\!\lim_{\sigma\rightarrow\infty}\!\frac{\int\! d_{i}\!\left(\!\frac{\rho(\sigma;\omega)}{\sigma}\mathcal{N}(\frac{d_{i}}{\sigma};0,\sigma^{2})\!+\!(1\!-\!\rho(\sigma;\omega))\delta(d_{i})\!\right)\!\mathcal{N}(d_{i};R,\Sigma)dd_{i}}
{\!\int\! \left(\!\frac{\rho(\sigma;\omega)}{\sigma}\mathcal{N}(\frac{d_{i}}{\sigma};0,\sigma^{2})\!+\!(1\!-\!\rho(\sigma;\omega))\delta(d_{i})\!\right)\!\mathcal{N}(d_{i};R,\Sigma)dd_{i}},\\
\end{split}
\end{equation}
and
\begin{equation}\label{ap4}
\begin{split}
f_{v_{i}}(\Sigma,R)
\!=\!\!\lim_{\sigma\rightarrow\infty}\!\!\frac{\int\!\! d_{i}\!\left(\!\frac{\rho(\sigma;\omega)}{\sigma}\mathcal{N}(\frac{d_{i}}{\sigma};0,\sigma^{2})\!\!+\!\!(1\!-\!\rho(\sigma;\omega))\delta(d_{i})\!\right)\!\mathcal{N}(d_{i};R,\Sigma)dd_{i}}
{\!\!\int\! \left(\!\frac{\rho(\sigma;\omega)}{\sigma}\mathcal{N}(\frac{d_{i}}{\sigma};0,\sigma^{2})\!\!+\!\!(1-\rho(\sigma;\omega))\delta(d_{i})\!\right)\!\mathcal{N}(d_{i};R,\Sigma)dd_{i}}-f_{a_{i}}^{2}(\Sigma,R).\\
\end{split}
\end{equation}
To yield the desired results, (\ref{ap3}) and (\ref{ap4}) can be further simplified as follows:
\begin{equation}\label{ap5}
\begin{split}
f_{a_{i}}(\Sigma,R)
&\overset{(a)}{=}\!\lim_{\sigma\rightarrow\infty}\!\frac{\!\!\int\!\! d_{i}\frac{\rho(\sigma;\omega)}{\sigma}\mathcal{N}(\frac{d_{i}}{\sigma};0,\sigma^{2})\mathcal{N}(d_{i};R,\Sigma)dd_{i}}{\!\!\int\!\! \frac{\rho(\sigma;\omega)}{\sigma}\mathcal{N}(\frac{d_{i}}{\sigma};0,\sigma^{2})\mathcal{N}(d_{i};R,\Sigma)dd_{i}\!\!+\!\!(1\!\!-\!\!\rho(\sigma;\omega))\mathcal{N}\!(0;R,\Sigma)\!}\\
&\overset{(b)}{=}\frac{\int d_{i}\mathcal{N}(d_{i};R,\Sigma)dd_{i}}{\int \mathcal{N}(d_{i};R,\Sigma)dd_{i}+\sqrt{2\pi\Sigma}e^{\omega}\mathcal{N}(0;R,\Sigma)}\\
&\overset{(c)}{=}\frac{R}{1+e^{\omega-\frac{R^{2}}{2\Sigma}}},\\
\end{split}
\end{equation}
and
\begin{equation}\label{ap6}
\begin{split}
f_{v_{i}}(\Sigma,R)
&\overset{(b)}{=}\frac{\int d^{2}_{i}\mathcal{N}(d_{i};R,\Sigma)dd_{i}-R^{2}+R^{2}}{\int \mathcal{N}(d_{i};R,\Sigma)dd_{i}+\sqrt{2\pi\Sigma}e^{\omega}\mathcal{N}(0;R,\Sigma)}\!-\!\left(\frac{R}{1+e^{-\frac{R^{2}}{2\Sigma}+\omega}}\right)^{2}\\
&\overset{(d)}{=}\frac{1}{1+e^{\omega-\frac{R^{2}}{2\Sigma}}}\left(\Sigma+\frac{R^{2}}{1+e^{\frac{R^{2}}{2\Sigma}-\omega}}\right),\\
\end{split}
\end{equation}
where (a) holds because $\int f(x)\delta(x)dx=f(0)$, and (b) can be established based on $\rho(\sigma;\omega)=\frac{\sigma}{\sigma+\mathcal{N}(0;0,\sigma^{2})\sqrt{2\pi\Sigma}e^{\omega}}$. Since $\int d_{i}\mathcal{N}(d_{i};R,\Sigma)dd_{i}=R $ and $\int \mathcal{N}(d_{i};R,\Sigma)dd_{i}=1$ according to the probability theory, (c) can be obtained. Besides $\int d^{2}_{i}\mathcal{N}(d_{i};R,\Sigma)dd_{i}-R^{2}= \Sigma$ explains (d).

\vspace{0.7em}
\emph{Case} \uppercase\expandafter{\romannumeral2}: In this case ($i\leq MN$), $\mathbf{h}$ follows uniform distribution as expressed in (\ref{u1}). The conditional mean and variance estimators for this case can be deduced as follows:
\begin{equation}\label{ap7}
\begin{split}
&f_{a_{i}}(\Sigma,R)=\frac{\int h_{i}\mathcal{N}(h_{i};R,\Sigma)dh_{i}}{\int \mathcal{N}(h_{i};R,\Sigma)dh_{i}}\overset{(a)}{=}R,\\
&f_{v_{i}}(\Sigma,R)=\frac{\int h_{i}^{2}\mathcal{N}(h_{i};R,\Sigma)dh_{i}}{\int \mathcal{N}(h_{i};R,\Sigma) dh_{i}}-R^{2}\overset{(b)}{=}\Sigma,\\
\end{split}
\end{equation}
where (a) and (b) hold because $\int h_{i}\mathcal{N}(h_{i};R,\Sigma)dh_{i}\!\!=\!\!R $ and $\int \mathcal{N}(h_{i};R,\Sigma)dd_{i}=1$, as well as $\int h_{i}^{2}\mathcal{N}(h_{i};R,\Sigma)dh_{i}-R^{2}= \Sigma$.

\vspace{0.7em}
\emph{Case} \uppercase\expandafter{\romannumeral3}: In this case ($i\leq MN$), $\mathbf{h}$ follows sparse Gaussian distribution as expressed in (\ref{g1}). The similar conclusion can be obtained for this case as follows:
\begin{equation}\label{ap8}
\begin{split}
&f_{a_{i}}(\Sigma,R)=\frac{\int h_{i}P(h_{i})\mathcal{N}(h_{i};R,\Sigma)dh_{i}}{\int P(h_{i})\mathcal{N}(h_{i};R,\Sigma)dh_{i}},\\
\end{split}
\end{equation}
and
\begin{equation}\label{ap9}
\begin{split}
&f_{v_{i}}(\Sigma,R)=\frac{\int h^{2}_{b,i}P(h_{i})\mathcal{N}(h_{i};R,\Sigma)dh_{i}}{\int P(h_{i})\mathcal{N}(h_{i};R,\Sigma)dh_{i}}-f_{a_{i}}^{2}(\Sigma,R).\\
\end{split}
\end{equation}
Substituting (\ref{g1}) into (\ref{ap8}) and (\ref{ap9}), we obtain
\begin{equation}\label{ap10}
\begin{split}
f_{a_{i}}(\Sigma,R)
=\frac{\int h_{i}\lambda\mathcal{N}(h_{i};a,v)\mathcal{N}(h_{i};R,\Sigma)dh_{i}}{\int \lambda\mathcal{N}(h_{i};a,v)\mathcal{N}(h_{i};R,\Sigma)dh_{i}+(1-\lambda)\mathcal{N}(0;R,\Sigma)},\\
\end{split}
\end{equation}
and
\begin{equation}\label{ap100}
\begin{split}
f_{v_{i}}(\Sigma,R)
=\frac{\int h^{2}_{b,i}\lambda\mathcal{N}(h_{i};a,v)\mathcal{N}(h_{i};R,\Sigma)dh_{i}}{\int \lambda\mathcal{N}(h_{i};a,v)\mathcal{N}(h_{i};R,\Sigma)dh_{i}+(1-\lambda)\mathcal{N}(0;R,\Sigma)}
-f_{a_{i}}^{2}(\Sigma,R),\\
\end{split}
\end{equation}
respectively.
Since
\begin{equation}
\mathcal{N}(x;a,v\!) \mathcal{N}(x;R,\Sigma)
=\mathcal{N}(0;a-R,v+\Sigma) \mathcal{N}\Bigg(x;\frac{\frac{a}{v}+\frac{R}{\Sigma}}{\frac{1}{v}+\frac{1}{\Sigma}},\frac{1}{\frac{1}{v}+\frac{1}{\Sigma}}\Bigg),
\end{equation}
we have
\begin{equation}\label{ap101}
\begin{split}
f_{a_{i}}(\Sigma,R)
=\frac{\int h_{i}\lambda\mathcal{N}(0;a-R,v+\Sigma)\mathcal{N}(h_{i};\frac{\frac{a}{v}+\frac{R}{\Sigma}}{\frac{1}{v}+\frac{1}{\Sigma}},\frac{1}{\frac{1}{v}+\frac{1}{\Sigma}})dh_{i}}{ \lambda\mathcal{N}(0;a-R,v+\Sigma)+(1-\lambda)\mathcal{N}(0;R,\Sigma)}\\
\end{split}
\end{equation}
and
\begin{equation}\label{ap11}
\begin{split}
f_{v_{i}}(\Sigma,R)
=\frac{\int h^{2}_{b,i}\lambda\mathcal{N}(0;a-R,v+\Sigma)\mathcal{N}(h_{i};\frac{\frac{a}{v}+\frac{R}{\Sigma}}{\frac{1}{v}+\frac{1}{\Sigma}},\frac{1}{\frac{1}{v}+\frac{1}{\Sigma}})dh_{i}}{\lambda\mathcal{N}(0;a-R,v+\Sigma)+(1-\lambda)\mathcal{N}(0;R,\Sigma)}
-f_{a_{i}}^{2}(\Sigma,R),\\
\end{split}
\end{equation}
respectively.
For further simplification, we obtain
\begin{equation}\label{ap102}
\begin{split}
f_{a_{i}}(\Sigma,R)
=\frac{\lambda\mathcal{N}(0;a-R,v+\Sigma)\frac{\frac{a}{v}+\frac{R}{\Sigma}}{\frac{1}{v}+\frac{1}{\Sigma}}}{\lambda\mathcal{N}(0;a-R,v+\Sigma)+(1-\lambda)\mathcal{N}(0;R,\Sigma)}
=\frac{\frac{a}{v}+\frac{R}{\Sigma}}{(\frac{1}{v}+\frac{1}{\Sigma})(1+\eta)}
\end{split}
\end{equation}
and
\begin{equation}\label{ap11}
\begin{split}
f_{v_{i}}(\Sigma,R)&
=\frac{\lambda\mathcal{N}(0;a-R,v+\Sigma)(\frac{1}{\frac{1}{v}+\frac{1}{\Sigma}}+(\frac{\frac{a}{v}+\frac{R}{\Sigma}}{\frac{1}{v}+\frac{1}{\Sigma}})^2)  }{\lambda\mathcal{N}(0;a-R,v+\Sigma)+(1-\lambda)\mathcal{N}(0;R,\Sigma)}
-f_{a_{i}}^{2}(\Sigma,R)
=\frac{\left(\frac{1}{\frac{1}{v}+\frac{1}{\Sigma}}+(\frac{\frac{a}{v}+\frac{R}{\Sigma}}{\frac{1}{v}+\frac{1}{\Sigma}})^2\frac{\eta}{1+\eta}\right)}{1+\eta},
\end{split}
\end{equation}
where $\eta=\frac{1-\lambda}{\lambda}\frac{\mathcal{N}(0;R,\Sigma)}{\mathcal{N}(0;a-R,v+\Sigma)}=\frac{1-\lambda}{\lambda}\sqrt{\frac{v+\Sigma}{\Sigma}}e^{\frac{(a-R)^2}{2(v+\Sigma)}-\frac{R^{2}}{2\Sigma}}$.

\section{}
\vspace{-3mm}
We derive the EM updates of sparsity rate $\lambda$, mean $a$, and variance $v$. We only prove the update of variance $v$ because similar arguments can be applied to sparsity rate $\lambda$ and mean $a$.

To find the maximum likelihood value of $v$, we have to let the first derivation of the sum $\sum_{i=1}^{MN} \hat{E}\{\ln P(h_{i};\mathbf{q}^{t})|\mathbf{r};\mathbf{q}^{t}\}$ equal to zero, which can be alternatively written as
\begin{equation}\label{c2}
\sum_{i=1}^{MN} \int_{h_{i}}P_{h_{i}|\mathbf{r}}(h_{i}|\mathbf{r};\mathbf{q}^{t})\frac{d}{dv}\ln P(h_{i};\mathbf{q}^{t})dh_{i}=0,
\end{equation}
where $P_{h_{i}|\mathbf{r}}(h_{i}|\mathbf{r};\mathbf{q}^{t})$ is posterior marginal pdf. From (\ref{uuu}), we obtain
\begin{equation}\label{cc1}
P_{h_{i}|\mathbf{r}}(h_{i}|\mathbf{r};\mathbf{q}^{t})=\frac{P(h_{i}) \mathcal{N}(h_{i};\mathbf{W}_{i}\mathbf{h},\Delta)}{\int_{h_{i}} P(h_{i}) \mathcal{N}(h_{i};\mathbf{W}_{i}\mathbf{h},\Delta)dh_{i}}.
\end{equation}
For further simplification, we have
\begin{equation}\label{cc2}
\begin{split}
P_{h_{i}|\mathbf{r}}(h_{i}|\mathbf{r};\mathbf{q}^{t})
&=\frac{(1-\lambda)\mathcal{N}(0;\!\mathbf{W}_{i}\mathbf{h},\Delta)\delta(h_{i})}{(1\!-\!\lambda)\mathcal{N}(0;\!\mathbf{W}_{i}\mathbf{h},\Delta)\!\!+\!\!\lambda  \mathcal{N}\!(0;\!\mathbf{W}_{i}\mathbf{h}\!-\!a,\Delta\!+\!v)}\\
&+\frac{\lambda \mathcal{N}\!(0;\!\mathbf{W}_{i}\mathbf{h}\!-\!a,\Delta\!+\!v)\mathcal{N}\!(h_{i};\frac{\frac{a}{v}+\frac{\mathbf{W}_{i}\mathbf{h}}{\Delta}}{\frac{1}{\mathbf{W}_{i}\mathbf{h}}+\frac{1}{\Delta}},\frac{1}{\frac{1}{\mathbf{W}_{i}\mathbf{h}}+\frac{1}{\Delta}})}{(1\!-\!\lambda)\mathcal{N}(0;\!\mathbf{W}_{i}\mathbf{h},\Delta)\!\!+\!\!\lambda  \mathcal{N}\!(0;\!\mathbf{W}_{i}\mathbf{h}\!-\!a,\Delta\!+\!v)}.
\end{split}
\end{equation}
Let
\begin{equation}\label{cc3}
\gamma_{i} \triangleq \frac{\frac{a}{v}+\frac{\mathbf{W}_{i}\mathbf{h}}{\Delta}}{\frac{1}{\mathbf{W}_{i}\mathbf{h}}+\frac{1}{\Delta}},
\qquad \nu_{i}  \triangleq \frac{1}{\frac{1}{\mathbf{W}_{i}\mathbf{h}}+\frac{1}{\Delta}},
\end{equation}
and
\begin{equation}\label{cc4}
\pi_{i} \triangleq \frac{\lambda \mathcal{N}\!(0;\!\mathbf{W}_{i}\mathbf{h}\!-\!a,\Delta\!+\!v)}{(1\!-\!\lambda)\mathcal{N}(0;\!\mathbf{W}_{i}\mathbf{h},\Delta)+\lambda  \mathcal{N}\!(0;\!\mathbf{W}_{i}\mathbf{h}\!-\!a,\Delta\!+\!v)};
\end{equation}
then
\begin{equation}\label{cc5}
1-\pi_{i} \triangleq \frac{(1-\lambda)\mathcal{N}(0;\!\mathbf{W}_{i}\mathbf{h},\Delta)}{(1\!-\!\lambda)\mathcal{N}(0;\!\mathbf{W}_{i}\mathbf{h},\Delta)+\lambda  \mathcal{N}\!(0;\!\mathbf{W}_{i}\mathbf{h}\!-\!a,\Delta\!+\!v)}.
\end{equation}
Plugging (\ref{cc3})-(\ref{cc5}) into (\ref{cc2}), we obtain
\begin{equation}\label{cc6}
P_{h_{i}|\mathbf{r}}(h_{i}|\mathbf{r};\mathbf{q}^{t}) = \pi_{i}\mathcal{N}(h_{i};\gamma_{i},\nu_{i}) + (1-\pi_{i})\delta(h_{i}).
\end{equation}
Plugging the marginal probability distribution function
\begin{equation}\label{cc7}
P(h_{i};\mathbf{q}^{t})=\lambda\mathcal{N}(h_{i};a,v)+(1-\lambda)\delta(h_{i})
\end{equation}
 into $d\ln P(h_{i};\mathbf{q}^{t})/d\lambda$, we obtain
\begin{equation}\label{c3}
\begin{split}
\frac{d}{dv}\ln P(h_{i};\mathbf{q}^{t})
=\frac{d}{dv}\ln \left(\lambda\frac{1}{\sqrt{2\pi v}}\exp\{-\frac{(h_{i}-a)^{2}}{2v}\}
+(1-\lambda)\delta(h_{i})\right).\\
\end{split}
\end{equation}
Then (\ref{c3}) can be simplified to
\begin{equation}\label{c33}
\begin{split}
\frac{d}{dv}\ln P(h_{i};\mathbf{q}^{t})
=\frac{-\frac{1}{2}v^{-1}\lambda\mathcal{N}(h_{i};a,v)+\frac{(h_{i}-a)^{2}}{2}v^{-2}\lambda\mathcal{N}(h_{i};a,v)}{(1-\lambda)\delta(h_{i})+\lambda\mathcal{N}(h_{i};a,v)}
=\frac{\frac{1}{2}\lambda\mathcal{N}(h_{i};a,v)(\frac{(h_{i}-a)^{2}}{v^{2}}-\frac{1}{v})}{(1-\lambda)\delta(h_{i})+\lambda\mathcal{N}(h_{i};a,v)},\\
\end{split}
\end{equation}
because $\delta(h_{i})\!\!=\!\!0$ when $h_{i}\!\!\neq\!\!0$ and $\delta(h_{b,i})\!\!\rightarrow\!\!\infty$ when $h_{b,i}\!\!=\!\!0$. Thus, we have
\begin{equation}\label{c333}
\begin{split}
&\frac{d}{dv}\ln P(h_{b,i};\mathbf{q}^{t})=\begin{cases}
\frac{1}{2}(\frac{(h_{b,i}-a)^{2}}{v^{2}}-\frac{1}{v}) &\mbox{if $h_{b,i}\neq0$,}\\
0 &\mbox{if $h_{i}=0$.}
\end{cases}
\end{split}
\end{equation}
Plugging (\ref{c333}) into (\ref{c2}) yields
\begin{equation}\label{c4}
\sum_{i=1}^{MN} \int_{h_{i}\neq0}P_{h_{i}|\mathbf{r}}(h_{i}|\mathbf{r};\mathbf{q}^{t})(\frac{(h_{i}-a)^{2}}{v^{2}}-\frac{1}{v})dh_{i}=0.
\end{equation}
We simplify (\ref{c4}) to
\begin{equation}\label{c5}
\begin{split}
\sum_{i=1}^{MN} \int_{h_{i}\neq0}P_{h_{i}|\mathbf{r}}(h_{i}|\mathbf{r};\mathbf{q}^{t})\frac{(h_{i}-a)^{2}}{v^{2}}dh_{i}
=\sum_{i=1}^{MN} \int_{h_{i}\neq0}P_{h_{i}|\mathbf{r}}(h_{i}|\mathbf{r};\mathbf{q}^{t})\frac{1}{v}dh_{i},
\end{split}
\end{equation}
and then  (\ref{c5}) can be rewritten to
\begin{equation}\label{c6}
\begin{split}
\sum_{i=1}^{MN} \int_{h_{i}\neq0}P_{h_{i}|\mathbf{r}}(h_{i}|\mathbf{r};\mathbf{q}^{t})(h_{i}-a)^{2}dh_{i}=v\sum_{i=1}^{MN} \int_{h_{i}\neq0}P_{h_{i}|\mathbf{r}}(h_{i}|\mathbf{r};\mathbf{q}^{t})dh_{i}.
\end{split}
\end{equation}
Now, we can observe that
\begin{equation}\label{c7}
v=\frac{\sum_{i=1}^{MN} \int_{h_{i}\neq0}P_{h_{i}|\mathbf{r}}(h_{i}|\mathbf{r};\mathbf{q}^{t})(h_{i}-a)^{2}dh_{i}}{\sum_{i=1}^{MN} \int_{h_{i}\neq0}P_{h_{i}|\mathbf{r}}(h_{i}|\mathbf{r};\mathbf{q}^{t})dh_{i}}.
\end{equation}
With the knowledge of $P_{h_{i}|\mathbf{r}}(h_{i}|\mathbf{r};\mathbf{q}^{t})$ from (\ref{cc6}) and dependent quantities $\pi_{i},\gamma_{i},\nu_{i}$, we substitute $\pi_{i},\gamma_{i},\nu_{i}$ into the denominator in (\ref{c7}) and simplify to obtain
\begin{equation}\label{c8}
\begin{split}
&\sum_{i=1}^{MN} \int_{h_{i}\neq0}P_{h_{i}|\mathbf{r}}(h_{i}|\mathbf{r};\mathbf{q}^{t})(h_{i}-a)^{2}dh_{i}
=\sum_{i=1}^{MN}\pi_{i}\int_{h_{i}\neq0}\mathcal{N}(h_{i};\gamma_{i},\nu_{i})(h_{i}-a)^{2}dh_{i}.\\
\end{split}
\end{equation}
Then, we apply perfect square expression to yield
\begin{equation}\label{c88}
\begin{split}
&\sum_{i=1}^{MN} \int_{h_{i}\neq0}P_{h_{i}|\mathbf{r}}(h_{i}|\mathbf{r};\mathbf{q}^{t})(h_{i}-a)^{2}dh_{i}\\
&=\!\sum_{i=1}^{MN}\pi_{i}\!\int_{h_{i}\neq0}\!\!\mathcal{N}\!(h_{i};\gamma_{i},\nu_{i})h_{i}^{2}dh_{i}
\!+\!\sum_{i=1}^{MN}a^{2}\pi_{i}\!\int_{h_{i}\neq0}\!\!\mathcal{N}\!(h_{i};\gamma_{i},\nu_{i})dh_{i}
-2\sum_{i=1}^{MN}a\pi_{i}\!\int_{h_{i}\neq0}\!\!\mathcal{N}\!(h_{i};\gamma_{i},\nu_{i})h_{i}dh_{i},\\
\end{split}
\end{equation}
according to the probability theory, and we can easily show that
\begin{equation}\label{c888}
\begin{split}
\sum_{i=1}^{MN} \int_{h_{i}\neq0}P_{h_{i}|\mathbf{r}}(h_{i}|\mathbf{r};\mathbf{q}^{t})(h_{i}-a)^{2}dh_{i}=\sum_{i=1}^{MN}\pi_{i}(\nu_{i}+\gamma_{i}^{2}+a^{2}-2a\gamma_{i})=\sum_{i=1}^{MN}\pi_{i}(\nu_{i}+(\gamma_{i}-a)^{2}).
\end{split}
\end{equation}
Thus, by substituting (\ref{c888}) into (\ref{c7}), $v$ can be easily calculated as follows:
\begin{equation}\label{c9}
v=\frac{\sum_{i=1}^{MN}\pi_{i}(\nu_{i}+(\gamma_{i}-a)^{2})}{\sum_{i=1}^{MN}\pi_{i}}.
\end{equation}

From similar derivation, we can finally obtain
the EM update of $\lambda$, $a$, and $v$ as follows:
\begin{align}
\lambda^{t+1}&=\frac{1}{MN}\sum_{i=1}^{MN}\pi_{i}, \label{a6} \\
a^{t+1}&=\frac{1}{MN\lambda^{t+1}}\sum_{i=1}^{MN}\pi_{i}\gamma_{i},\label{b10} \\
v^{t+1}&=\frac{1}{MN\lambda^{t+1}}\sum_{i=1}^{MN}\pi_{i}(\nu_{i}+(\gamma_{i}-a^{t+1})^{2}). \label{c10}
\end{align}
The preceding equations are easily computed by plugging the expression of $\pi_{i},\gamma_{i},\nu_{i}$.

\end{appendices}

\ifCLASSOPTIONcaptionsoff
  \newpage
\fi





\begin{thebibliography}{1}

\bibitem{mmWave1} F. Giannetti, M. Luise, and R. Reggiannini, ``Mobile and personal communications in 60 GHz band: A survey," \emph{Wirelesss Pers. Commun.}, vol. 10, pp. 207-243, Jul. 1999.
\bibitem{mmWave2} H. Xu, V. Kukshya, and T. S. Rappaport, ``Spatial and temporal characteristics of 60 GHz indoor channel," \emph{IEEE J. Sel. Areas Commun.}, vol. 20, no. 3, pp. 620-630, Apr. 2002.
\bibitem{mmWave3} R. Daniels and R. W. Heath Jr, ``60 GHz wireless communications: Emerging requirements and design recommendations," \emph{IEEE Veh. Technol. Mag.}, vol. 2, no. 3, pp. 41-50, Sep. 2007.
\bibitem{overview} R. W. Heath Jr, N. G. Prelcic, S. Rangan, W. Roh, and A. Sayeed, ``An overview of signal processing techniques for millimeter wave MIMO systems," \emph{IEEE J. Sel. Topics Signal Process.}, vol. 10, no. 3, pp. 436-453, Apr. 2016.

\bibitem{mmWaveMIMO1} S. Singh, R. Mudumbai, and U. Madhow, ``Interference analysis for highly directional 60-GHz mesh networks: The case for rethinking medium access control," \emph{IEEE/ACM Trans. Netw.}, vol. 19, no. 5, pp. 1513-1527, Oct. 2011.

\bibitem{mmWaveMIMO2} E. Torkildson, U. Madhow, and M. Rodwell, ``Indoor millimeter wave MIMO: Feasibility and performance," \emph{IEEE Trans. Wireless Commun.}, vol. 10, no. 12, pp. 4150-4160, Dec. 2011.

\bibitem{hard1} A. Pyattaev, K. Johnsson, S. Andreev, and Y. Koucheryavy, ``Communication
challenges in high-density deployments of wearable wireless
devices," \emph{IEEE Wireless Commun.}, vol. 22, no. 1, pp. 12-18, Mar. 2015.

\bibitem{hy0} A. Alkhateeb, M. Jianhua, N. Gonzalez-Prelcic, and R. W. Heath Jr,
``MIMO precoding and combining solutions for millimeter-wave systems,"
\emph{IEEE Commun. Mag.}, vol. 52, no. 12, pp. 122-131, Dec. 2014.

\bibitem{hy1} O. El Ayach, S. Rajagopal, S. Abu-Surra, Z. Pi, and R. W. Heath Jr,
``Spatially sparse precoding in millimeter wave MIMO systems," \emph{IEEE
J. Sel. Areas Commun.}, vol. 13, no. 3, pp. 1499-1513, Mar. 2014.

\bibitem{hy2} A. Alkhateeb, O. El Ayach, G. Leus, and R. W. Heath Jr, ``Channel
estimation and hybrid precoding for millimeter wave cellular systems,"
\emph{IEEE J. Sel. Topics Signal Process.}, vol. 8, no. 5, pp. 831-846, Oct
2014.

\bibitem{hy3} S. Han, C. -L. I, Z. Xu, and C. Rowell, ``Large-scale antenna systems
with hybrid analog and digital beamforming for millimeter wave 5G,"
\emph{IEEE Commun. Mag.}, vol. 53, no. 1, pp. 186-194, Jan. 2015.

\bibitem{1bit1}  J. Singh, O. Dabeer, and U. Madhow, ``On the limits of communication with low-precision analog-to-digital conversion at the receiver", \emph{IEEE Trans. Commun.}, vol. 57, no. 12, pp. 3629-3639, Dec. 2009.


\bibitem{lens1} W. Rotman and R. Turner, ``Wide-angle microwave lens for line source applications," \emph{IEEE Trans. Antennas Propag.}, vol. 11, no. 6, pp. 623-632, Nov. 1963.
\bibitem{lens2} D. T. McGrath, ``Planar three-dimensional constrained lenses," \emph{IEEE Trans. Antennas Propag.}, vol. 34, no. 1, pp. 46-50, Jan. 1986.
\bibitem{lens3} Z. Popovic and A. Mortazawi, ``Quasi-optical transmit/receive front ends," \emph{IEEE Trans. Microw. Theory Tech.}, vol. 48, pp. 1964-1975, Nov. 1998.
\bibitem{lens4} G. Godi, R. Sauleau, L. Le Coq, and D. Thouroude, ``Design and optimization of three dimensional integrated lens antennas with genetic algorithm," \emph{IEEE Trans. Antennas Propag.}, vol 55, no. 3, pp. 770-775, Mar. 2007.
\bibitem{lens5} R. Sauleau and B. Bares, ``A complete procedure for the design and optimization of arbitrarily-shaped integrated lens antennas," \emph{IEEE Trans. Antennas Propag.}, vol 54, no. 4, pp.122-133, Apr. 2006.


\bibitem{beamspace} J. Brady, N. Behdad, and A. M. Sayeed, ``Beamspace MIMO for millimeter-wave communications: System architecture, modeling, analysis, and measurements," \emph{IEEE Trans. Antennas Propag.}, vol. 61, no. 7, pp. 3814-3827, Jul. 2013.
\bibitem{1} Y. Zeng, and R. Zhang, ``Millimeter wave MIMO with lens antenna array: A new path division multiplexing paradigm," \emph{IEEE Trans. Commun.}, vol. 64, no. 4, pp. 1557-1571, Apr. 2016.
\bibitem{7} Y. Zeng, L. Yang, and R. Zhang, ``Multi-user millimeter wave MIMO with full-dimensional lens antenna array," \emph{arXiv preprint arXiv:1611.06008}, 2016.


\bibitem{111} A. Alkhateeb, G. Leus, and R. W. Heath Jr, ``Compressed sensing based  multi-user millimeter wave systems: How many measurements are needed?" in \emph{Proc. IEEE Int. Conf. Acoustics, Speech and Sig. Process. (ICASSP)}, Brisbane, Australia, Apr. 2015, pp. 2909-2913.
\bibitem{LASSO} R. Tibshirani, ``Regression shrinkage and selection via the lasso," \emph{Journal of Royal Statistical Society. Series B (Methodological)}, pp. 267-288, 1996.
\bibitem{OMP} T. Tony Cai and L. Wang, ``Orthogonal matching pursuit for sparse signal recovery with noise," \emph{IEEE Trans. Inf. Theory}, vol. 57, no. 7, pp. 4680-4688, Jun. 2011.
\bibitem{cs1} W. U. Bajwa, J. Haupt, A. M. Sayeed, and R. Nowak, ``Compressed channel sensing: A new approach to estimating sparse multipath channels," in \emph{Proc. IEEE}, vol. 98, no. 6, pp. 1058-1076, Jun. 2010.
\bibitem{cs2} X. Rao and V. K. N. Lau, ``Distributed compressive CSIT estimation and feedback for FDD multi-user massive MIMO systems," \emph{IEEE Trans. Signal Process.}, vol. 62, no. 12, pp. 3261-3271, Jun. 2014.
\bibitem{cs3} Y. Shi, J. Zhang, and K. B. Letaief, ``CSI overhead reduction with stochastic beamforming for cloud radio access networks," in \emph{Proc. IEEE Int. Conf. Commun. (ICC)}, Sydney, N. S. W, Australia, Jun. 2014, pp. 5154-5159.
\bibitem{cs4} S. Nguyen and A. Ghrayeb, ``Compressive sensing-based channel estimation for massive multiuser MIMO systems," in \emph{Proc. IEEE Wireless Commun. Networking Conf. (WCNC)}, Shanghai, China, Apr. 2013, pp. 2890-2895.
\bibitem{cs5} S. Nguyen and A. Ghrayeb, ``Precoding for multicell MIMO systems with compressive rank-q channel approximation," in \emph{Proc. IEEE Annual Int. Symposium on Personal, Indoor, and Mobile Radio Commun. (PIMRC)}, London, U. K., Sep. 2013, pp. 1227-1232.


\bibitem{jin} C. K. Wen, S. Jin, K. K. Wong, J. C. Chen, and P. Ting, ``Channel estimation for massive MIMO using Gaussian-mixture bayesian learning," \emph{IEEE Trans. Wireless Commun.}, vol. 14, no. 3, pp. 1356-1368, Mar. 2015.

\bibitem{3} L. Dai, X. Gao, S. Han, C.-L. I, and X. Wang, ``Beamspace channel estimation for millimeter-wave massive MIMO systems with lens antenna array," in \emph{Proc. IEEE/CIC Int. Conf. Computer and Commun. (ICCC)}, Chengdu, China, Jul. 2016, pp. 1-6.

\bibitem{bethe} F. Krzakala, A. Manoel, E. W. Tramel, and L. Zdeborov$\acute{a}$, ``Variational
free energies for compressed sensing," in \emph{Proc. IEEE Int. Symposium on Inf. Theory (ISIT)}, Honolulu, U.S.A., Jul. 2014,
pp. 1499-1503.


\bibitem{5} J. Barbier, E. W. Tramel, and F. Krzakala, ``SCAMPI: A robust approximate message-passing framework for compressive imaging," \emph{Journal of Phys. Conf. Series}, vol. 699, Mar. 2016.

\bibitem{6} M. Borgerding, P. Schniter, and S. Rangan, ``Generalized approximate message passing for cosparse analysis compressive sensing," in \emph{Proc. IEEE Int. Conf. Acoustics, Speech and Sig. Process. (ICASSP)}, Brisbane, Australia, Apr. 2015, pp. 3756-3760.



\bibitem{em3} A. Dempster, N. M. Laird, and D. B. Rubin, ``Maximum-likelihood from incomplete data via the EM algorithm," \emph{Journal of Royal Statistical Society.}, vol. 39, pp. 1-17, 1977.
\bibitem{em1} J. -P. Vila and P. Schniter, ``Expectation-maximization bernoulli-gaussian approximation message passing," in \emph{Proc. Asilomar Conf. Signals, Systs., Comput.}, Pacific Grove, CA, USA, Nov. 2011, pp. 799-803.
\bibitem{em2} J. -P. Vila and P. Schniter, ``Expectation-maximization gaussian-mixture approximate message passing," in \emph{Proc. Conf. Inf. Sci. Syst.(CISS)}, Princeton, NJ, USA, Mar. 2012, pp. 1-6.



\end{thebibliography}
\end{document}